\documentclass[10pt, a4paper]{article} %[twoside,12pt,a4paper]{article}

\usepackage{graphicx}
\usepackage{times,mathptmx}
\usepackage{fancyhdr}
\usepackage{indentfirst}
\usepackage{color}

\usepackage{url}
\usepackage{amsmath}
\usepackage{caption}
\usepackage{subcaption}
\makeatletter
\renewcommand\Large{\@setfontsize\Large{16.00}{20.00}}
\renewcommand \thesection {\@arabic\c@section.}
\renewcommand\thesubsection   {\thesection\@arabic\c@subsection.}
\renewcommand\thesubsubsection{\thesubsection\@arabic\c@subsubsection.}
\renewcommand\section{\@startsection {section}{1}{\z@}%
				   {-3.5ex \@plus -1ex \@minus -.2ex}%
				   {2.3ex \@plus.2ex}%
				   {\normalfont\normalsize\bfseries}}
\renewcommand\subsection{\@startsection{subsection}{2}{\z@}%
				     {-2ex\@plus -1ex \@minus -.2ex}%
				     {.1ex \@plus .2ex}%
				     {\normalfont\normalsize\itshape}}
\renewcommand\subsubsection{\@startsection{subsubsection}{3}{\z@}%
				     {-2ex\@plus -1ex \@minus -.2ex}%
				     {.1ex \@plus .2ex}%
				     {\normalfont\normalsize\itshape}}

\long\def\@makecaption#1#2{%
  \vskip\abovecaptionskip
  \sbox\@tempboxa{#1 #2}%
  \ifdim \wd\@tempboxa >\hsize
    #1 #2\par
  \else
    \global \@minipagefalse
    \hb@xt@\hsize{\hfil\box\@tempboxa\hfil}%
  \fi
  \vskip\belowcaptionskip}
\makeatother
\begin{document}

%%title
{\noindent
\Large\bfseries
Towards robust coupled field induced Josephson junctions %%\newline
 %%%in complex electromagnetic environments
\\}

%%author
{\noindent
Krzysztof Pomorski$^{1,2,3}$ 
}

%%affiliation
{\noindent
\itshape
\\
%%$^1\!$
1: University College Dublin, School of Computer Science, \\ Belfield, Dublin 4, Eircode: D04 V1W8, Ireland \\ \\
2: Nagoya University, Quantum Engineering Department, Furo-cho, Chikusa-ku, \newline Nagoya 464-8603, Japan \\ \\
3: Quantum Hardware Systems: \url{www.quantumhardwaresystems.com} 

%Eircode:  D04 V1W8
 %%255 Shimo-Okubo, Sakura-ku, Saitama 338-8570, Japan\\
%$^2\!$
%Department of Electrical and Computer Eng., Yokohama National University, 79-5 Tokiwadai, Hodogaya, Yokohama 240-8501, Japan
}

%%email
{\noindent \\
E-mail: kdvpomorski@gmail.com$^{1,2}$ %t\_saitama@super.ees.saitama-u.ac.jp
}
\begin{center}
\rule[6pt]{6.1in}{1.4pt}
\end{center}
\vspace*{-.3in}

\section*{Abstract}
The concept of coupled robust field induced Josephson junctions placed in complex electromagnetic environments is presented. Such structures are made by polarization of superconducting nanostructures by magnetic fields. The methodology of modeling of structures and possible implementations is introduced. The presented scheme is expected to implement and describe both classical and quantum computers with use of Josephson junction and artificial evolution.
In such cases one can obtain unexpected new topologies of circuits that can contribute in enhancement of known and used circuit schemes. Furthermore one can obtain the non-invasive detectors of moving charged particles from robust field induced Josephson junctions which are illustrated by examples. 

%%%The concept of coupled robust field induced Josephson junctions placed in complex electromagnetic environments is presented. The methodology of modeling of structures and possible implementations is introduced. The presented scheme is expected to implement and describe both classical and quantum computer with use of Josephson junction and artificial evolution.
%%%In such case one can obtain unexpected new topologies of circuits that can contribute in enhancement of known and used circuit schemes.
%%An introduction for preparing manuscripts for the proceedings of Superconducting SFQ VLSI Workshop (SSV Workshop) 2016 held on August 2$^\mathrm{nd}$
%%and 3$^\mathrm{rd}$, 2016, at Yokohama National University (Yokohama, Japan) is given in the present paper.
%%This paper is written in the format to be used for all manuscripts and illustrates the final appearance for the proceedings.
%%It is very important to follow these instructions to both ensure that your manuscript is quickly accepted, and to present consistent and professional looking workshop proceedings publication.

\section{Research motivation in relation to superconducting electronics}

The Josephson effect \cite{Josephson} allows to use unique properties of superconducting
macroscopic quantum effect. Great progress was achieved in theoretical understanding of this phenomena and in implementation of Josephson junctions in
various physical systems as in superconductors, superfluids, Bose-Einstein condensate or polaritons. From perspective of current development of electronics
the most promising is the usage of low temperature s-wave superconductors although highly correlated materials and majorana fermions are also important topics of fundamental research. Likharev proposed the usage of fluxons \cite{Likharev} as representation of logical bit 1 and until now it seems to be the most optimal implementation of classical information in superconductor that gives the base for
development of Rapid Single Quantum Flux (RSFQ) electronics. The greatest integration of Josephson junctions was achieved indeed in RSFQ electronics \cite{SQFMicroprocessor}, \cite{Mukhanov}. The development of complex superconducting qubit circuits \cite{Nori} is also promising but not so dynamical as it is the case of RSFQ electronics. In implementation of Josephson junction circuits most common is the use of standard
topology of superconductor and insulators or non-superconducting materials
which makes design and implementation more easy to be achieved from technological and simulation point of view. This results in assigning concept of simple
washboard potential to particular Josephson junctions of rather simple topology as presented in \cite{Nori}. Presence of simple form of washboard potentials simplifies description of
the system from computational point of view. Nevertheless the simplicity of
washboard potential does not exploit many physical phenomena that could take
place and which might be technically useful. The purpose of this
work is to draw and motivate new research direction that can take into account
broad class of washboard potentials with various complexities and topologies. This shall have its relevance in implementation of both classical and quantum computer in superconductor which requires use of system with many Josephson junctions. It should also have its relevance in studying various properties of
condensates from fundamental point of view and also is related to science of complexity. Thus it is highly interdisciplinary research as it is shown in Fig.\ref{fig6q}. \newline

In the next section the definition of field induced Josephson junction will be given and later it will be generalized to various classes of physical systems. Section 3 gives mathematical insight into FIJJ from the point of view of 1-dimensional Ginzburg-Landau theory. Section 4 describes the use of robust field induced Josephson junction for non-invasive detection of moving charge again using GL theory. Section 5 is pointing the use of Bogoliubov-de Gennes theory basing on previously obtained GL solutions. Finally the possible experimental setups and
schemes expressing various class of topologies for robust field induced Josephson junctions are drawn and new possible future directions of research are briefly discussed.

%%All papers should be written in English.
%%The title should be followed by the name(s) of the author(s), the institution and its address.
%%The outline of the paper should be set out using appropriate headings, i.e. Abstract, 1. Introduction, 2. Experimental, 3. Results, 4. Discussion, and 5. Conclusions.
%%In order to publish the proceedings of papers in time, items listed below must
%%{\bfseries \color{red}{ONLINE}} no later than the deadline date of
%%{\bfseries \color{red}{MONDAY, JULY 4$^\mathbf{TH}$, 2016}}.

%%\begin{itemize}
%%\item Upload an electric version of the camera-ready manuscript (Word file, or {\LaTeX} file and EPS files of figures)\\
%%  via the SSV 2016 website at
%%\begin{center}
%%  \color{blue}{\underline{http://www.super.nuqe.nagoya-u.ac.jp/SSV2016/}}
%%\end{center}
%%\end{itemize}

\section{Concept of Field Induced Josephson junctions (FIJJs) and its generalizations}

The first Josephson junction system was obtained for superconductor-insulator- %\newline
superconductor with thin insulator barrier \cite{Josephson}. In such case one of the most important
parameters is the probability of tunneling from one superconductor into another
for wavepackets of certain energy. Later it was shown that broad class of constrictions (narrowings) in superconductor can generate the effect similar to that
observed in tunneling Josephson junction. In such a way, weak-link Josephson
junctions were found as the system with higher transmission probability than
the case of tunneling structure. However one should not only limit in usage of
physical constrains encoded in physical lattice \cite{BTK} (and hence boundary conditions)
to obtain Josephson effect. One can use the external magnetic or electric fields
in order to modulate magnitude of superconducting order parameter (SCOP).
First idea of usage of magnetic field in modulating the superconducing order parameter was coming from Clinton \cite{Clinton} although the concept of topological defect
in superconducting order parameter is old. Clinton has placed the ferromagnetic
strip on the top of superconduciting strip separated with insulating barrier. In
such case the fringe field coming from ferromagnet was used to diminish superconducting order parameter locally. Further idea of similar kind was developed
by L.Gomez and A.Maeda \cite{Maeda} and relied on direct evaporating of ferromagnetic material on the top
of superconducting strip. In such case one deals with diffusion of spin polarized particles from ferromagnet into superconductor and with penetration of
magnetic field into superconductor. Various possible effects can occur as induction of triplet phase in s-wave superconductor. Due to many possible physical
phenomena and complex mathematical picture in two or three dimensions as
pointed in \cite{Pomorski} it is rather recommended to study at first one dimensional superconducting strip polarized with external source magnetic field as it is depicted
in four subfigures of Fig.\ref{fig1q}(situation Fig.\ref{fig1q}a is nearly equivalent to Fig.\ref{fig1q}c and case of  Fig.\ref{fig1q}b
can be approximated by Fig.\ref{fig1q}d in the limit of long normal strip). It could be equivalent to Clinton/Johnson system of ferromagnet placed on the superconducting with
presence of insulating barrier that blocks spin polarized electron diffusion from
ferromagnet into superconductor. We assume that thickness of superconductor
is up to two superconducting coherence lengths so we can neglect the variation
of superconducting order parameter across superconductor and assume that it
is constant. In first approximation we will neglect the occurrence of quantum
phase slips \cite{Nazarov} in superconducting nanowire.

Transport properties of structures can be determined when we combine both Ginzburg-Landau and
Bogoliubov-de Genne formalism. It should be pointed that we should specify the asymptotic states at first so SCOP and vector potential would be constant.
In general case the polarizing cables topology and distribution of polarizing current is variational problem when we try to approximate Fig.\ref{fig1q}c by \ref{fig1q}a or \ref{fig1q}d by \ref{fig1q}b.
In general case we have two classes of polarizing cables generating magnetic field that polarizes superconductor (closed loops with $A_x$ and $A_y$ vector potential components and open infinite cable generating $A_z$ vector potential) and
we assume that they are not affected by superconductor. The Bogoliubov-de Gennes (BdGe) wavepacket propagating via superconductor is affected by the
presence of non-uniform vector potential and hence is partly scattered when it propagates from ($x=-\infty$ to $x=+\infty$). The shape of polarizing loops is not fixed. Quite similar situation could occur when we have continuously deformed superconducting wire (quasi-one dimensionality is preserved and this allows the usage of non-linear Ordinary Differential Equations (ODE) instead of non-liner Partial Differential Equations(PDE) ) as it is the case of Fig.\ref{fig2q}a. In similar fashion we can deform
the polarizing cables as well which can result in amorphic and robust but field induced Josephson junction depicted in Fig.\ref{fig2qb}. Continuing this type of reasoning we can obtain two (or more) coupling in inductive way FIJJs as it is depicted in Fig.\ref{fig2qc}. Most general
scheme is given by Fig.\ref{fig2qd} when one can deal with net of coupled robust field induced Josephson junctions embedded in network of polarizing cables of complex topology.
\section{Usage of vector potential for generation of field-induced Josephson junction}
%\subsection{Polarizing current perpendicular to superconducting nanowire (PC1)}

\begin{figure}[htb!]
    \centering
 \includegraphics[width=4.0in]{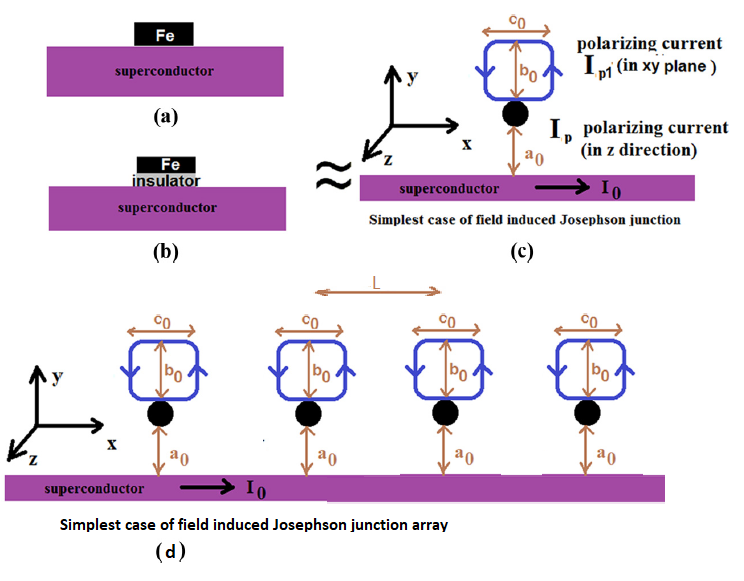} %%%{JJ1.png}
 \caption{Field Induced Josephson Junction (FIJJ) approximated by external current driven polarizing loop as given in \cite{Compel}. We recognize two basic geometrical configurations as polarizing current in parallel (PC1) and perpendicular to superconducting nanowire (PC2). }
 \label{fig:FIJJ}
\end{figure}

We begin considerations with the physical situation of polarizing cable that is perpendicular to superconducting nanowire (CP1) as depicted in Fig.\ref{fig:FIJJ}. We have given the following phase imprint on superconducting nanowire due to Aharonov-Bohm effect
\begin{eqnarray}
\alpha(x')=\frac{\mu_0}{4\pi}\frac{ j_p \Delta z}{\sqrt{a_0^2+(x')^2}}\frac{e}{c \hbar},
\end{eqnarray}
where $\Delta z$ is thickness of superconducting nanowire and $j_p$ is value of intensity of polarizing current density. Therefore the macroscopic wavefunction also known as SCOP (Superconducting Order Parameter) is given as
\begin{eqnarray}
\psi(x')=|\psi(x')|e^{i\alpha(x')}=|\psi(x')|e^{i \frac{\mu_0}{4\pi}\frac{ j_p \Delta z}{\sqrt{a_0^2+(x')^2}}\frac{e}{c \hbar}}=|\psi(x')|e^{i \frac{\mu_0}{4\pi}\frac{I_p \Delta z}{\sqrt{a_0^2+(x')^2}}\frac{e}{c \hbar}},
\end{eqnarray}
 and is referring to the GL equation
 \begin{equation}
 \alpha_s(x)\psi(x)+\beta(x)|\psi(x)|^2\psi(x)+\frac{1}{2m}(\frac{\hbar}{i}\frac{d}{dx}-\frac{2e}{c}A_x(x))^2\psi(x)=0
 \end{equation}
that is equivalent to
 \begin{equation}
 \alpha_s(x)\psi(x)+\beta(x)|\psi(x)|^2\psi(x)+\frac{1}{2m}(-\hbar^2\frac{d^2}{dx^2}+\frac{4e^2}{c^2}A_x(x)^2+i\hbar\frac{2e}{c}(\frac{d}{dx}A_x(x))+i\hbar\frac{4e}{c}A_x(x)\frac{d}{dx})\psi(x)=0 \end{equation}
and we obtain
\begin{eqnarray}
\frac{d}{dx'}\psi(x')=e^{i\alpha(x')}[\frac{d}{dx'}|\psi(x')|+i|\psi(x')|(\frac{d}{dx'}\alpha(x'))], \nonumber \\
\frac{d^2}{dx'^2}\psi(x')=e^{i\alpha(x')}[\frac{d^2}{dx'^2}|\psi(x')|+i(\frac{d}{dx'}|\psi(x')|)(\frac{d}{dx'}\alpha(x'))+i|\psi(x')|(\frac{d^2}{dx'^2}\alpha(x'))+ \nonumber \\
+i(\frac{d}{dx'}\alpha(x'))(\frac{d}{dx'}|\psi(x')|)-|\psi(x')|(\frac{d}{dx'}\alpha(x'))^2]
\end{eqnarray}
We can extract only imaginary part of GL equation and we obtain
\begin{eqnarray}
\frac{-\hbar^2}{2m}[i(\frac{d}{dx'}|\psi(x')|)(\frac{d}{dx'}\alpha(x'))+i|\psi(x')|(\frac{d^2}{dx'^2}\alpha(x'))+i(\frac{d}{dx'}\alpha(x'))(\frac{d}{dx'}|\psi(x')|)]+\frac{1}{2m}i\hbar\frac{2e}{c}(\frac{d}{dx}A_x(x))|\psi(x')|= \nonumber \\
=-\frac{1}{2m}i\hbar\frac{4e}{c}A_x(x')(\frac{d}{dx'}|\psi(x')|)
\end{eqnarray}
Last equation is of first order ODE for real valued functions $|\psi(x')|=\sqrt{n(x')}$ and $\alpha(x')$ and we obtain
\begin{eqnarray}
\frac{-\hbar c}{4m e}[(\frac{d}{dx'}|\psi(x')|)(\frac{d}{dx'}\alpha(x'))+|\psi(x')|(\frac{d^2}{dx'^2}\alpha(x'))+(\frac{d}{dx'}\alpha(x'))(\frac{d}{dx'}|\psi(x')|)]+\frac{1}{2m}(\frac{d}{dx}A_x(x))|\psi(x')|= \nonumber \\
=\frac{1}{2m}(-2A_x(x')(\frac{d}{dx'}|\psi(x')|)).
\end{eqnarray}
After rearrangement
\begin{eqnarray*}
[\frac{-\hbar c}{e}[(\frac{d}{dx'}\alpha(x'))]+2A_x(x')](\frac{d}{dx'}|\psi(x')|) %+ \nonumber \\
=[-(\frac{d}{dx}A_x(x))+\frac{\hbar c}{2 e}(\frac{d^2}{dx'^2}\alpha(x'))]|\psi(x')| \nonumber \\
%=-2A_x(x')(\frac{d}{dx'}|\psi(x')|).
\end{eqnarray*}
and it brings
\begin{eqnarray*}
(\frac{d|\psi(x')|}{|\psi(x')|}) %+ \nonumber \\
=\frac{[-(\frac{d}{dx}A_x(x'))+\frac{\hbar c}{2e}(\frac{d^2}{dx'^2}\alpha(x'))]}{[\frac{-\hbar c}{e}[(\frac{d}{dx'}\alpha(x'))]+2A_x(x')]}dx' \nonumber \\
%=-2A_x(x')(\frac{d}{dx'}|\psi(x')|).
\end{eqnarray*}
that can be integrated
\begin{eqnarray*}
ln(\frac{|\psi(x_2)|}{|\psi(x_1)|}) %+ \nonumber \\
=\int_{x1}^{x2}\frac{[-(\frac{d}{dx}A_x(x'))+\frac{\hbar c}{2 e}(\frac{d^2}{dx'^2}\alpha(x'))]}{[\frac{-\hbar c}{ e}[(\frac{d}{dx'}\alpha(x'))]+2A_x(x')]}dx' \nonumber \\
%=-2A_x(x')(\frac{d}{dx'}|\psi(x')|).
\end{eqnarray*}

and it implies for perpendicular polarizing cable ($A_x=0$) the relation to be as
\begin{equation}
|\psi(x_2,t)|=|\psi(x_1,t)| e^{-\frac{1}{2}\int_{x1}^{x2}dx' \Bigg[\frac{\frac{d^2 \alpha_x(x')}{dx'^2}}{\frac{d\alpha_x(x')}{dx'}} \Bigg] }=|\psi(x_1)| e^{-\frac{1}{2}\int_{x1}^{x2}dx' \Bigg[\frac{\frac{d^2}{dx'^2}\frac{1}{\sqrt{a_0^2+(x')^2}}}{\frac{d}{dx'}\frac{1}{\sqrt{a_0^2+(x')^2}}} \Bigg] }= \nonumber \\
\frac{\sqrt{x_1}
   \left(a_0^2+x_2^2\right)^{\frac{3}{4}}}{\sqrt{x_2}
   \left(a_0^2+x_1^2\right)^{\frac{3}{4}}}|\psi(x_1)|
%=e^{-\frac{1}{2}\int_{x1}^{x2}(\frac{1}{x'} - \frac{3x}{(a_0^2 + x'^2)})dx'}= \nonumber \\
%=e^
\end{equation}
We can also extract only real value part of Ginzburg-Landau equation in 1 dimension subjected to arbitrary potential vector polarization and we obtain
 \begin{equation}
 \alpha_s(x)|\psi(x)|+\beta(x)|\psi(x)|^3+\frac{1}{2m}(-\hbar^2\frac{d^2}{dx^2}|\psi(x)|+(\hbar\frac{d}{dx}\alpha(x))^2|\psi(x)|+\frac{4e^2}{c^2}(A_x^2+A_y^2+A_z^2)|\psi(x)|-\hbar\frac{4e}{c}A_x(x)\frac{d\alpha(x)}{dx}|\psi(x)|)=0 \end{equation}
 and it leads to the equation
 \begin{equation}
 \Bigg[\alpha_s(x)+(\hbar\frac{d}{dx}\alpha(x))^2+\frac{1}{2m}\frac{4e^2}{c^2}(A_x^2+A_y^2+A_z^2)-\hbar\frac{4e}{c}A_x(x)\frac{d\alpha(x)}{dx}\Bigg]|\psi(x)|+\beta(x)|\psi(x)|^3-\frac{\hbar^2}{2m}\frac{d^2}{dx^2}|\psi(x)|=0, \end{equation}
 so coefficient $\alpha_s(x)$ describing the state of superconductor without presence of external magnetic field (vector potential) is replaced with coefficient $\alpha_{1s}(x)$ that is less negative (or less superconductive) than $\alpha_s(x)$, where
  \begin{eqnarray}
 \alpha_{1s}(x)=\Bigg[\alpha_s(x)+(\hbar\frac{d}{dx}\alpha(x))^2+\frac{1}{2m}\frac{4e^2}{c^2}(A_x^2+A_y^2+A_z^2)-\hbar\frac{4e}{c}A_x(x)\frac{d\alpha(x)}{dx}\Bigg], \nonumber \\ \alpha_{1s}(x)|\psi(x)|+\beta(x)|\psi(x)|^3-\frac{\hbar^2}{2m}\frac{d^2}{dx^2}|\psi(x)|=0.
  \end{eqnarray}
%%We notice the equivalence of the chain of equations
%% \begin{eqnarray}
%% \alpha_{0}|\psi(x)|+\beta|\psi(x)|^3-\frac{\hbar^2}{2m}\frac{d^2}{dx^2}|\psi(x)|=0, \nonumber \\
%% \alpha_0|\psi(x)|\frac{d}{dx}|\psi(x)|+\beta|\psi(x)|^3\frac{d}{dx}|\psi(x)|-\frac{\hbar^2}{2m}(\frac{d^2}{dx^2}|\psi(x)|)\frac{d}{dx}|\psi(x)|=0, \nonumber \\
%%  \frac{d}{dx} (\frac{\alpha_0}{2}|\psi(x)|^2 +\frac{\beta}{4}|\psi(x)|^4 -\frac{\hbar^2}{4m}(\frac{d}{dx}|\psi(x)| )^2 )= \frac{d}{dx}F, F=constans, \nonumber \\
%%  (\frac{\alpha_0}{2}|\psi(x)|^2 +\frac{\beta}{4}|\psi(x)|^4 -\frac{\hbar^2}{4m}(\frac{d}{dx}|\psi(x)| )^2 )= F
%%  \end{eqnarray}
%\subsection{Polarizing current paralell to superconducting nanowire (PC2)}
In particular having closed rectangular-shape polarizing loop (PC1) as depicted in  Fig.\ref{fig:FIJJ} we have two non-zero components of vector potential $A_x$ and $A_y$ and geometrical constrains given by ($a_0, b_0, c_0$), so finally we obtain formulas:
\begin{eqnarray}
A_x(x,x_2=+\frac{c_0}{2},x_1=-\frac{c_0}{2})=\frac{\mu_0 I_p}{4\pi}\Bigg[ \log \Bigg[
   \left(\sqrt{\frac{\sqrt{a_0^2+(\frac{c_0}{2}+x)^2}+(\frac{c_0}{2}+x)}{\sqrt{a_0^2
   +(\frac{c_0}{2}+x)^2}-(\frac{c_0}{2}+x)}}
   \sqrt{\frac{\sqrt{a_0^2+(\frac{c_0}{2}-x)^2}+(\frac{c_0}{2}-x)}{\sqrt{a_0^2+(\frac{c_0}{2}-x)^2}-(\frac{c_0}{2}-x)}}\right)\times \nonumber \\
   \times
   \left(\sqrt{\frac{\sqrt{(a_0+b_0)^2+(\frac{c_0}{2}+x)^2}-(\frac{c_0}{2}+x)}{\sqrt{(a_0+b_0)^2
   +(\frac{c_0}{2}+x)^2}-(\frac{c_0}{2}+x)}}
   \sqrt{\frac{\sqrt{(a_0+b_0)^2+(\frac{c_0}{2}-x)^2}+(\frac{c_0}{2}-x)}{\sqrt{(a_0+b_0)^2+(\frac{c_0}{2}-x)^2}-(\frac{c_0}{2}-x)}}\right)
    \Bigg] \Bigg], %\nonumber \\
\end{eqnarray}
\begin{eqnarray}
   A_y(x=x_s,y=y_s=0,x_2=\frac{c_0}{2},x_1=-\frac{c_0}{2},y_2=a_0+b_0,y_1=a_0)= \nonumber \\
   = \frac{\mu_0 I_p}{4\pi}\Bigg[  \log \Bigg[
   \left( \sqrt{\frac{\sqrt{(\frac{c_0}{2}+x)^2+(a_0)^2}+a_0}{\sqrt{(\frac{c_0}{2}+x)^2+(a_0)^2}-a_0}}
   \sqrt{\frac{\sqrt{(\frac{c_0}{2}+x)^2+(a_0+b_0)^2}-(a_0+b_0)}{\sqrt{(\frac{c_0}{2}+x)^2+(a_0+b_0)^2}+a_0+b_0}}\right)  \times \nonumber \\
  \times
   \left(\sqrt{\frac{\sqrt{(\frac{c_0}{2}-x)^2+(a_0)^2}-a_0}{\sqrt{(\frac{c_0}{2}-x)^2+(a_0)^2}+a_0}}
   \sqrt{\frac{\sqrt{(\frac{c_0}{2}-x)^2+(a_0+b_0)^2}+(a_0+b_0)}{\sqrt{(\frac{c_0}{2}-x)^2+(a_0+b_0)^2}-(a_0+b_0)}}\right) \Bigg] \Bigg] ,
\end{eqnarray}
\begin{eqnarray}
   \alpha(x)=(\frac{2e}{\hbar c}([\int_{x0}^{x}dx'A_x(x',+\frac{c_0}{2},-\frac{c_0}{2})]+\Delta y A_y(x,0,\frac{c_0}{2},-\frac{c_0}{2})) , \nonumber \\
   \frac{d}{dx}\alpha(x)=(\frac{2e}{\hbar c}(A_x(x,+\frac{c_0}{2},-\frac{c_0}{2})+\Delta y \frac{d}{dx}A_y(x,0,\frac{c_0}{2},-\frac{c_0}{2})), \nonumber \\
\end{eqnarray}
and consequently we obtain the following equation for superconducting order parameter from Ginzburg-Landau formalism for Field Induced Josephson junction in the following form
\begin{eqnarray}
 \Bigg[\alpha_s(x)+(\hbar\frac{d}{dx}\alpha(x))^2+\frac{1}{2m}\frac{4e^2}{c^2}(A_x^2+A_y^2)-\hbar\frac{4e}{c}A_x(x)\frac{d\alpha(x)}{dx}\Bigg]|\psi(x)| = \nonumber \\
 = \Bigg[\alpha_s(x)+ %\nonumber \\
 (\hbar (\frac{2e}{\hbar c}(A_x(x,+\frac{c_0}{2},-\frac{c_0}{2})+\Delta y \frac{d}{dx}A_y(x,0,\frac{c_0}{2},-\frac{c_0}{2})))^2+\frac{1}{2m}\frac{4e^2}{c^2}(A_x(x,+\frac{c_0}{2},-\frac{c_0}{2})^2+ \nonumber \\ +A_y(x,+\frac{c_0}{2},-\frac{c_0}{2})^2) %% \nonumber \\
  -\hbar\frac{4e}{c}A_x(x)(\frac{2e}{\hbar c}(A_x(x,+\frac{c_0}{2},-\frac{c_0}{2})+\Delta y \frac{d}{dx}A_y(x,0,\frac{c_0}{2},-\frac{c_0}{2}))\Bigg]|\psi(x)|= \nonumber \\
  =-\beta(x)|\psi(x)|^3+\frac{\hbar^2}{2m}\frac{d^2}{dx^2}|\psi(x)|.
 %\alpha_{1s}(x)|\psi(x)|+\beta(x)|\psi(x)|^3-\frac{\hbar^2}{2m}\frac{d^2}{dx^2}|\psi(x)|=0.
    %-\frac{\mu_0 I_p}{4\pi}\Bigg[-\log
  % \left(\sqrt{\frac{\sqrt{(x_1-x)^2+(y_{1})^2}+y_1}{\sqrt{(x_1-x)^
  % 2+(y_1)^2}-y_1}\right)+ \nonumber \\
  % +\log \left(\sqrt{\frac{\sqrt{(x_{1}-x)^2+(y_{2})^2}+y_{2}}{\sqrt{(x_1-x)^2+
  % (y_{2})^2}-y_{2}}}\right)+ \nonumber \\
%   +\log \left(\sqrt{\frac{\sqrt{(x_2-x)^2+(y_1)^2}+y_{1}}{\sqrt{(x_{2}-x)^2
%   +(y_{1})^2}-y_{1}}}\right) %%%+ \nonumber \\
% -\log \left(\sqrt{\frac{\sqrt{(x_{2}-x)^2+(y_{2})^2}+y_{2}}{\sqrt{(x_{2}-x)^2 +(y_{2})^2}-y_{2}}}\right) \Bigg],
\end{eqnarray}
where $(x_1, x_2)$ and $(y_1, y_2)$ are describing the geometrical parameters of polarizing circuit, so $c_0=x_2-x_1$ and $b_0=y_2-y_1$.
We have electric current density given by Ginzburg-Landau theory that is brings strong constrain in quasi-one dimensional wire and is expressed by formula
\begin{eqnarray}
  constant=J=\frac{1}{2m}[\psi(x)^{\dag}\hat{p}_x\psi(x)-\psi(x)\hat{p}_x\psi(x)^{\dag}]=\frac{1}{2m}[\psi(x)^{\dag}(\frac{\hbar}{i}\frac{d}{dx}-\frac{2e}{c}A_x)\psi(x)-\psi(x)(\frac{\hbar}{i}\frac{d}{dx}-\frac{2e}{c}A_x)\psi(x)^{\dag}]= \nonumber \\
  =\frac{1}{2m}[|\psi(x)|e^{-i\alpha(x)}(\frac{\hbar}{i}\frac{d}{dx}-\frac{2e}{c}A_x)|\psi(x)|e^{i\alpha(x)}-|\psi(x)|e^{i\alpha(x)}(\frac{\hbar}{i}\frac{d}{dx}-\frac{2e}{c}A_x)|\psi(x)|e^{-i\alpha(x)}] =
  %|\psi|^2v_x 2e=
  \nonumber \\
 =
 \frac{2e}{m}(-\hbar \frac{d}{dx}\alpha(x)-\frac{2e}{c}A_x)|\psi(x)|^2=\nonumber \\ =-\frac{4e^2}{mc}((A_x(x,+\frac{c_0}{2},-\frac{c_0}{2})+\Delta y \frac{d}{dx}A_y(x,0,\frac{c_0}{2},-\frac{c_0}{2})))|\psi(x)|^2 % =|\psi(t)|^2v_x(t)m \\
  % &=&  \\
  % &=& 
\end{eqnarray}
Last equation implies that superconducting order parameter can be determined uniquely and in analytic way in the form
\begin{eqnarray}
  \sqrt{\frac{|J|}{\frac{4e^2}{mc}|(A_x(x,+\frac{c_0}{2},-\frac{c_0}{2})+\Delta y \frac{d}{dx}A_y(x,0,\frac{c_0}{2},-\frac{c_0}{2})))|}}=|\psi(x)| % =|\psi(t)|^2v_x(t)m \\
  % &=&  \\
  % &=& 
\end{eqnarray}
The observed current is the consequence of Meissner effect that tends to shield the presence of magnetic field and thus is tending to minimize the system energy. 
Let us assume that we have the infinite chain of polarizing rectangular loops with polarizing electric current $I_p$ that are creating one dimensional lattice so the 
system has translational symmetry. 
We have the following expression for vector potentials and $A_x(x,y,z)$ has the following form \newline
\begin{eqnarray}
A_x(x,y,z)=I_{p}\frac{\mu}{2\pi}\Bigg[-\tanh^{-1}\left(\frac{\frac{c_{0}}{2}+L-x}{\sqrt{(a_{0}+b_{0}-y)^2+\left(-\frac{c_{0}}{2}-L+x\right)^2+z^2}}\right)+ \tanh^{-1}\left(\frac{-\frac{c_{0}}{2}+L-x}{\sqrt{(a_{0}+b_{0}-y)^2+\left(\frac{c_{0}}{2}-L+x\right)^2+z^2}}\right) \nonumber \\
-\tanh^{-1}\left(\frac{\frac{c_{0}}{2}-L-x}{\sqrt{(a_{0}+b_{0}-y)^2+\left(-\frac{c_{0}}{2}+L+x\right)^2+z^2}}\right)+  %\nonumber \\
\tanh^{-1}\left(\frac{-\frac{c_{0}}{2}+L-x}{\sqrt{(a_{0}+b_{0}-y)^2+\left(\frac{c_{0}}{2}+L+x\right)^2+z^2}}\right)  \nonumber \\
-\tanh^{-1}\left(\frac{\frac{c_{0}}{2}-x}{\sqrt{(a_{0}+b_{0}-y)^2+\left(x-\frac{c_{0}}{2}\right)^2+z^2}}\right)+  %\nonumber \\
\tanh^{-1}\left(\frac{-\frac{c_{0}}{2}-x}{\sqrt{(a_{0}+b_{0}-y)^2+\left(\frac{\text{c0}}{2}+x\right)^2+z^2}}\right)  \nonumber \\
+\tanh^{-1}\left(\frac{\frac{c_{0}}{2}+L-x}{\sqrt{(a_{0}-y)^2+\left(-\frac{c_{0}}{2}-L+x\right)^2+z^2}}\right)  %\nonumber \\
-\tanh^{-1}\left(\frac{-\frac{c_{0}}{2}+L-x}{\sqrt{(a_{0}-y)^2+\left(\frac{c_{0}}{2}-L+x\right)^2+z^2}}\right)  \nonumber \\
+\tanh^{-1}\left(\frac{\frac{c_{0}}{2}-L-x}{\sqrt{(a_{0}-y)^2+\left(-\frac{c_{0}}{2}+L+x\right)^2+z^2}}\right)  %\nonumber \\
-\tanh^{-1}\left(\frac{-\frac{c_{0}}{2}+L-x}{\sqrt{(a_{0}-y)^2+\left(\frac{c_{0}}{2}+L+x\right)^2+z^2}}\right)  \nonumber \\
+\tanh^{-1}\left(\frac{\frac{c_{0}}{2}-x}{\sqrt{(a_{0}-y)^2+\left(x-\frac{c_{0}}{2}\right)^2+z^2}}\right)  %\nonumber \\
-\tanh^{-1}\left(\frac{-\frac{c_{0}}{2}-x}{\sqrt{(a_{0}-y)^2+\left(\frac{c_{0}}{2}+x\right)^2+z^2}}\right) \Bigg]+  \nonumber \\
I_{sc} \Bigg[\left(\tanh ^{-1}\left(\frac{\frac{3 L}{2}-x}{\sqrt{\left(x-\frac{3 L}{2}\right)^2+y^2+z^2}}\right)  %%%%\nonumber \\
-\tanh^{-1}\left(\frac{-\frac{3 L}{2}-x}{\sqrt{\left(\frac{3 L}{2}+x\right)^2+y^2+z^2}}\right)\right)\Bigg]
\end{eqnarray}
and 
\begin{eqnarray}
A_y(x,y,z)=I_{p}\Bigg[ \tanh^{-1}\left(\frac{-a_{0}-b_{0}+y}{\sqrt{(-a_{0}-b_{0}+y)^2+\left(-\frac{c_{0}}{2}-L+x\right)^2+z^2}}\right) %\nonumber \\
+\tanh^{-1}\left(\frac{-a_{0}-b_{0}+y}{\sqrt{(-a_{0}-b_{0}+y)^2+\left(\frac{c_{0}}{2}-L+x\right)^2+z^2}}\right)  \nonumber \\
+\tanh^{-1}\left(\frac{-a_{0}-b_{0}+y}{\sqrt{(-a_{0}-b_{0}+y)^2+\left(-\frac{c_{0}}{2}+L+x\right)^2+z^2}}\right)  %\nonumber \\
+\tanh^{-1}\left(\frac{-a_{0}-b_{0}+y}{\sqrt{(-a_{0}-b_{0}+y)^2+\left(\frac{c_{0}}{2}+L+x\right)^2+z^2}}\right)  \nonumber \\
+\tanh^{-1}\left(\frac{-a_{0}-b_{0}+y}{\sqrt{(-a_{0}-b_{0}+y)^2+\left(x-\frac{c_{0}}{2}\right)^2+z^2}}\right)  %\nonumber \\
+\tanh^{-1}\left(\frac{-a_{0}-b_{0}+y}{\sqrt{(-a_{0}-b_{0}+y)^2+\left(\frac{c_{0}}{2}+x\right)^2+z^2}}\right)  \nonumber \\
-\tanh^{-1}\left(\frac{y-a_{0}}{\sqrt{(y-a_{0})^2+\left(-\frac{c_{0}}{2}-L+x\right)^2+z^2}}\right)  %\nonumber \\
-\tanh^{-1}\left(\frac{y-a_{0}}{\sqrt{(y-a_{0})^2+\left(\frac{c_{0}}{2}-L+x\right)^2+z^2}}\right)  \nonumber \\
-\tanh^{-1}\left(\frac{y-a_{0}}{\sqrt{(y-a_{0})^2+\left(-\frac{c_{0}}{2}+L+x\right)^2+z^2}}\right)  %\nonumber \\
-\tanh^{-1}\left(\frac{y-a_{0}}{\sqrt{(y-a_{0})^2+\left(\frac{c_{0}}{2}+L+x\right)^2+z^2}}\right)  \nonumber \\
-\tanh^{-1}\left(\frac{y-a_{0}}{\sqrt{(y-a_{0})^2+\left(x-\frac{c_{0}}{2}\right)^2+z^2}}\right)  %\nonumber \\
-\tanh ^{-1}\left(\frac{y-a_{0}}{\sqrt{(y-a_{0})^2+\left(\frac{c_{0}}{2}+x\right)^2+z^2}}\right) \Bigg]
\end{eqnarray}
In conducted considerations the superconducting nanowire coordinates was (x,y=0,z=0). 
The magnetic field present in the system is $B_x(x,y,z)=-\frac{d}{dz}A_y(x,y,z)$, $B_y(x,y,z)=\frac{d}{dx}A_z(x,y,z)-\frac{d}{dz}A_x(x,y,z)$ and \newline $B_z(x,y,z)=\frac{d}{dx}A_y(x,y,z)-\frac{d}{dy}A_x(x,y,z)$. The total magnetic energy present in the space for $x \in (-\frac{L}{2},\frac{L}{2})$, $y \in (-\infty,+\infty)$ and $z \in (-\infty,+\infty)$ is given by formula $U_E=\int_{-\frac{3}{2}L}^{{+\frac{3}{2}L}}\frac{1}{2\mu_0}\int_{-\infty}^{+\infty}dy\int_{-\infty}^{+\infty}dz(B_x(x,y,z)^2+B_y(x,y,z)^2+B_z(x,y,z)^2)$. Such considerations are done as the system has periodicity, so one has the same value of magnetic field if we are shifted by $x \rightarrow x \pm nL$. 
Here $I_p$ is the polarizing current with non-zero value and $I_{sc}$ is screening current. This screening current is tending to minimize the energy of magnetic field $U_E$.  
Value of $I_{sc}$ can be determined from the condition $\frac{d}{I_{sc}}U_E=$ that lead so the condition $\frac{d}{dI_{sc}}(B_x(x,y,z)^2+B_y(x,y,z)^2+B_z(x,y,z)^2)=0$. In such way value of screening current can be obtained analytically. The obtained results show that Field Induced Josephson junctions are varphi Josephson junctions with built-in phase difference. It simply means that ground state of the system is not when there is zero phase difference across Josephson junction. 
 It is important to refer to real values of superconducting coherence length for real materials as it is described in Table 1. Once we have determined the mathematical structure describing properties of FIJJ junctions in Ginzburg-Landau formalism we are moving towards the construction of non-invasive detector of moving charged particles from architecture of robust field induced Josephson junctions as it is described in the next section.
\section{Non-invasive superconducting detector of charged particles in time-dependent Ginzburg-Landau theory}
We assume that there is heavy charged particle with mass m and charge q that is moving along straight line and is affecting superconducting loop by time dependent electric and magnetic field and that superconducting loop response is too weak to affect the moving particle trajectory and dynamical state.
The moving particle is generating electrostatic charge and vector potential that is affecting superconducting loop. Electric current is generated in the superconducting loop. This electric current has two components that is superconducting non-dissipative current and current of quasiparticles that is dissipative.
The macroscopic wavefunction of superconductor has the following form as
\begin{equation}
\psi(r,\alpha,t)=\sqrt{n(\alpha,t)}e^{i\frac{2e}{\hbar c}\int_{0}^{\alpha}A_{\phi}(\alpha_1,t)rd\alpha_1}e^{i\frac{2e}{\hbar c}A_{r}(\alpha,t)\Delta R}
\end{equation}

\begin{figure}[hbt!]
\centering
\includegraphics[scale=0.3]{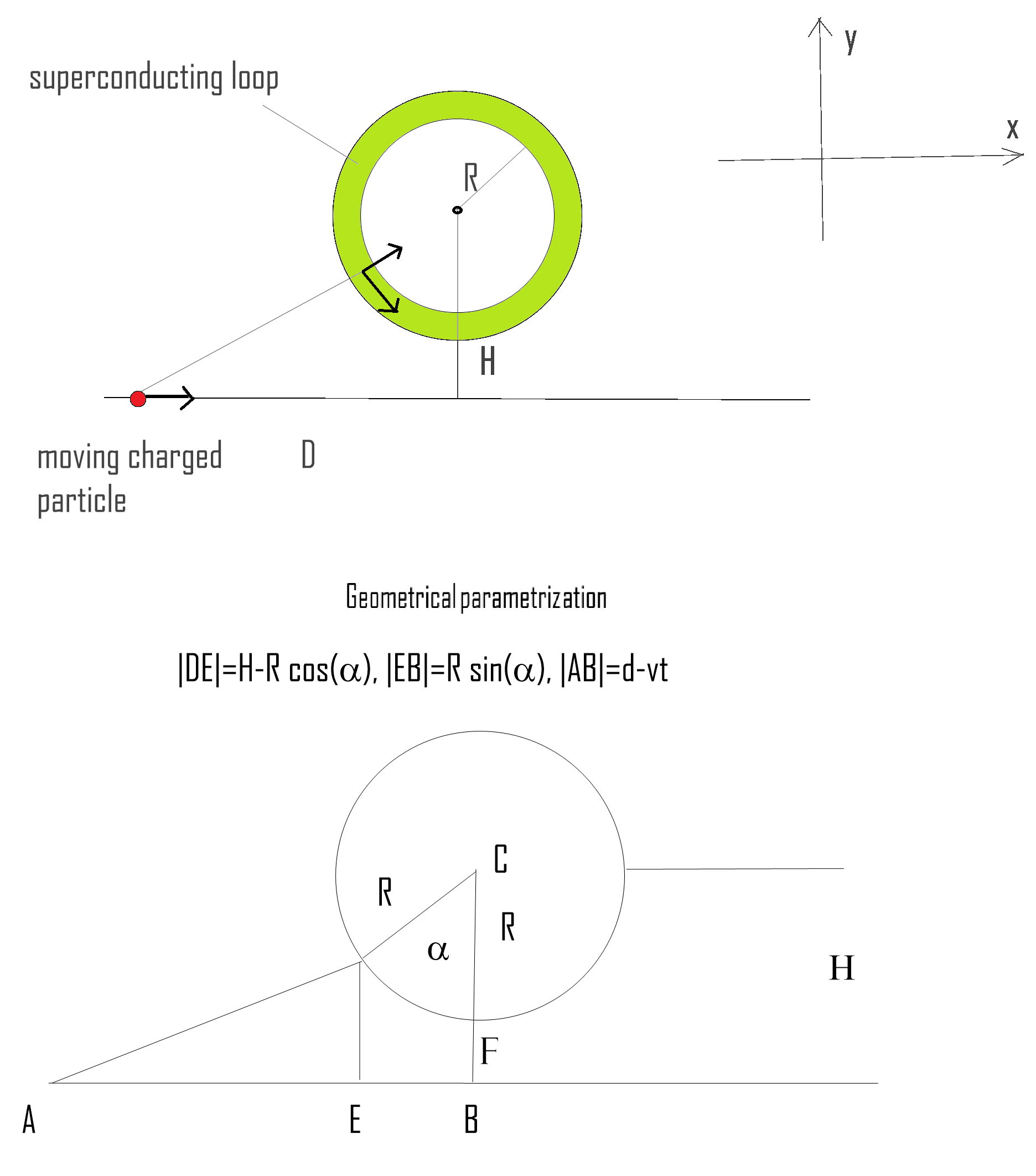} %%{Rys1qq.png} %{ScDetectorQQ.png}
\caption{Scheme of non-invasive superconducting detector of charged particles as the chain of closed superconducting loops. In depicted situation heavy charged particle is moving along the straight line with constant speed and is inducing time-dependent electric and magnetic fields acting on the superconducting loop. In more advanced considerations one needs to account for counteracting force from superconducting loop and acting on the charged particle that is deflecting its trajectory and thus is decreasing the impact of charged particle on the state of superconducting loop.  }
\end{figure}

We have two components of electric current that is normal and superconducting component of electric current. They can be expressed as

\begin{eqnarray}
\vec{J}=\vec{J}_{sc}+\vec{J}_{n}=\sigma_0(1-|\psi|^2) \vec{E} -(\frac{2e^2}{mc}\vec{A}+E_{\phi})|\psi(x)|^2, \sigma_0=\frac{q^2\tau}{m}.
\end{eqnarray}
Superconducting component of electric current is not subjected to dissipation and non-superconducting component of electric current
follows Ohm's law. Here $\tau$ is the relaxation time occurring in Drude model and it is recognizable that going to infinity is reproducing the superconducting state that is in very real way the highly ordered state. In particular we can have the formula for the electric current at the ring given as

\begin{eqnarray}
\vec{J}_{\phi}=\vec{J}_{sc,\phi}+\vec{J}_{n,\phi}=\sigma_0(1-|\psi|^2)(-\frac{d}{dt}\vec{A}_{\phi}-\vec{\nabla}_{\phi} V) -(\frac{2e^2}{mc}\vec{A}_{\phi}-2e\vec{E}_{\phi} )|\psi(x)|^2, \sigma_0=\frac{q^2\tau}{m}.
\end{eqnarray}
It is important to recognize that
\begin{eqnarray}
I(t)=-\sigma_0(1-|\psi(\alpha,r,t)|^2)(\frac{d}{dt}A_{\phi}(\alpha,r,t)+\frac{1}{r}\frac{d}{d\alpha} V(\alpha,r,t)) +(-\frac{2e^2}{mc}A_{\phi}(\alpha,r,t)-\frac{1}{r}\frac{d}{d\alpha} V(\alpha,r,t))|\psi(\alpha,r,t)|^2.
\end{eqnarray}
and that total current is non-depending on $\alpha$ and is only time-dependent.
The ratio of superconducting vs normal electric current at given point is given as
\begin{eqnarray}
\frac{J_{sc,\phi}}{J_{n,\phi}}=\frac{\sigma_0(1-|\psi|^2)(\frac{d}{dt}A_{\phi}+\frac{1}{r}\frac{d}{d\alpha} V)}{ +(\frac{2e}{mc}A_{\phi}+2e\frac{1}{r}\frac{d}{d\alpha} V)|\psi(x)|^2},
\end{eqnarray}

Therefore superconducting order parameter has to self-adjust in the way that right side of equation with all added terms does not depend on the angle $\alpha$. We obtain the following equation for superconducting order parameter dependence with angle $\alpha$ as
\begin{eqnarray}
\frac{I(t)+\sigma_0(\frac{d}{dt}A_{\phi}(\alpha,r,t)+\frac{1}{r}\frac{d}{d\alpha} V(\alpha,r,t))}{[\sigma_0(\frac{d}{dt}A_{\phi}(\alpha,r,t)+\frac{1}{r}\frac{d}{d\alpha} V(\alpha,r,t)) +\frac{2e^2}{mc}A_{\phi}(\alpha,r,t)]}=|\psi(\alpha,r,t)|^2.
\end{eqnarray}
Equivalently we have
\begin{eqnarray}
\sqrt{\frac{I(t)+\sigma_0(\frac{d}{dt}A_{\phi}(\alpha,r,t)+\frac{1}{r}\frac{d}{d\alpha} V(\alpha,r,t))}{[\sigma_0(\frac{d}{dt}A_{\phi}(\alpha,r,t)+\frac{1}{r}\frac{d}{d\alpha} V(\alpha,r,t)) +\frac{2e^2}{mc}A_{\phi}(\alpha,r,t)]}}=|\psi(\alpha,r,t)|. %=|\psi(\alpha,r,t)|_0+\delta |\psi(\alpha,r,t)|.
\end{eqnarray}
Now we are using the Maxwell equation for the case of closed loop of superconductor with hole inside.
We treat the system in quasi-one dimensional way
\begin{eqnarray}
\nabla \times \vec{E}= -\frac{d}{dt}\vec{B}
\end{eqnarray}
and it brings
\begin{eqnarray}
\int \vec{E} \cdot d\vec{x}= -\frac{d}{dt}\Phi_B(t)=-\pi r^2 \frac{d}{dt}I(t)-\frac{d}{dt}\Phi_{B,particle}
\end{eqnarray}
We have
\begin{eqnarray}
A_{\phi}=\frac{v_p q cos(\alpha)}{\sqrt{(d-r(sin(\alpha))-v_p t)^2+(H-r\cos(\alpha))^2}}, \nonumber \\
%\end{eqnarray}
%\begin{eqnarray}
A_{r}=\frac{v_p q sin(\alpha)}{\sqrt{(d-r(sin(\alpha))-v_p t)^2+(H-r\cos(\alpha))^2}},
\end{eqnarray}
and
\begin{eqnarray}
V=\frac{q^2}{\sqrt{(d-r sin(\alpha)-v_p t)^2+(H-r cos(\alpha))^2}}.
\end{eqnarray}
It is quite straightforward to obtain

\begin{eqnarray}
E_{\phi}(\alpha,r,t)=-\frac{1}{r}\frac{d}{d\alpha}V=-\frac{1}{r}\frac{d}{d\alpha} \frac{q^2}{\sqrt{(d-r sin(\alpha)-v_p t)^2+(H-r cos(\alpha))^2}}= \nonumber \\
=\frac{1}{r}\frac{q^2 (2 r \sin (\alpha ) (H-r \cos (\alpha ))-2 r \cos (\alpha ) (-r \sin (\alpha
   )+d-t v_p))}{2 \left((-r \sin (\alpha )+d-t v_p)^2+(H-r \cos (\alpha
   ))^2\right)^{3/2}}.
\end{eqnarray}
and

\begin{eqnarray}
\frac{d}{dt}A_{\phi}(\alpha,r,t)=
\frac{q v_p^2 \cos (\alpha ) (-r \sin (\alpha )+d-t v_p)}{\left((-r \sin (\alpha
   )+d-t v_p)^2+(H-r \cos (\alpha ))^2\right)^{3/2}}
\end{eqnarray}
Now it is time to compute the magnetic flux coming from external charged particle that is generating magnetic flux in the superconducting loop or circular constant radius.
At first we are computing the magnetic field generated by such particle and one arrives to the formula for non-zero z component of magnetic field under the assumption that x-y plane characterizes the position of superconducting loop as movement of the charged particle inducing changes to the superconducting loop. One arrives to the formula
\begin{eqnarray}
B_{z}(x_0,y_0,x_s,y_s)=\frac{\mu_0}{4\pi}v_p q \frac{y_s-y_0}{((x_s-x_0)^2+(y_s-y_0)^2)^{\frac{3}{2}}}=\frac{\mu_0}{4\pi}v_p q \frac{y_s-y_0}{((x_s-x_{p0}-v_pt)^2+(y_s-y_0)^2)^{\frac{3}{2}}},
\end{eqnarray}
where $(x_0,y_0)$ are characterizing the position of the charged particle and $(x_s,y_s)$ are parameterizing the geometrical point inside superconducting loop.
We have chosen the superconducting loop to be circular while in general case it can be of any shape.   The magnetic flux inside the superconducting loop coming from the charged particle is
\begin{eqnarray}
\phi_{B,particle}(t)=\frac{\mu_0}{4\pi}\int_{y1=-R}^{y2=-y1=R} \int_{x1(y1)}^{x2(y1)} B_{z}(x_0,y_0,x_s,y_s)dx_sdy_s= \nonumber \\
=\frac{\mu_0}{4\pi}\int_{y1=-R}^{y2=-y1=R} dy_s\int_{x1(y1)}^{x2(y1)} dx_s v_p q \frac{y_s-y_0}{((x_s-x_0)^2+(y_s-y_0)^2)^{\frac{3}{2}}}= \nonumber \\
=\frac{v_p q \mu_0}{4\pi}\int_{y1=-R}^{y2=-y1=R} dy_s [\frac{x_s-x_0}{y_s-y_0} \frac{1}{((x_s-x_0)^2+(y_s-y_0)^2)^{\frac{1}{2}}}]|_{x_{1s}=-\sqrt{R^2-y_s^2}}^{x_{2s}=\sqrt{R^2-y_s^2}} =\nonumber \\
=\frac{v_p q \mu_0}{4\pi}\int_{y1=-R}^{y_{2s}=-y_{1s}=R} dy_s [\frac{\sqrt{R^2-y_s^2}-x_{0p}-v_pt}{y_s-y_0} \frac{1}{((\sqrt{R^2-y_s^2}-x_{p0}-v_pt)^2+(y_s-y_0)^2)^{\frac{1}{2}}} \nonumber \\
+\frac{\sqrt{R^2-y_s^2}+x_{0p}+v_pt}{y_s-y_0} \frac{1}{((\sqrt{R^2-y_s^2}+x_{0p}+v_pt)^2+(y_s-y_0)^2)^{\frac{1}{2}}}].
\end{eqnarray}
The total magnetic flux inside the superconducting loop comes from moving charged particle and current flowing via superconducting loop.
Having the condition along the closed loop (CL) given as
\begin{eqnarray}
\int_{t0}^{t} dt' \int_{CL} \vec{E}(t') \cdot d\vec{x}= -\Phi_B=-g\pi r^2 I(t)-\Phi_{B,particle}(t)
\end{eqnarray}
we obtain

\begin{eqnarray}
\pi R^2 I(t)=-\int_{t0}^{t} dt' \int \vec{E} \cdot d\vec{x} \nonumber \\
-\frac{v_p q \mu_0}{4\pi}\int_{y1=-R}^{y_{2s}=-y_{1s}=R} dy_s [\frac{\sqrt{R^2-y_s^2}-x_{0p}-v_pt}{y_s-y_0} \frac{1}{((\sqrt{R^2-y_s^2}-x_{p0}-v_pt)^2+(y_s-y_0)^2)^{\frac{1}{2}}} \nonumber \\
+\frac{\sqrt{R^2-y_s^2}+x_{0p}+v_pt}{y_s-y_0} \frac{1}{((\sqrt{R^2-y_s^2}+x_{0p}+v_pt)^2+(y_s-y_0)^2)^{\frac{1}{2}}}]= \nonumber \\
=-\frac{1}{c}\int_{0}^{2\pi} R A_{\phi}\cdot d\phi -\int_{t0}^{t} dt' \int_{0}^{2\pi} d\phi R E_{\phi}+ \nonumber \\
-\frac{v_p q \mu_0}{4\pi}\int_{y1=-R}^{y_{2s}=-y_{1s}=R} dy_s [\frac{\sqrt{R^2-y_s^2}-x_{0p}-v_pt}{y_s-y_0} \frac{1}{((\sqrt{R^2-y_s^2}-x_{p0}-v_pt)^2+(y_s-y_0)^2)^{\frac{1}{2}}} \nonumber \\
+\frac{\sqrt{R^2-y_s^2}+x_{0p}+v_pt}{y_s-y_0} \frac{1}{((\sqrt{R^2-y_s^2}+x_{0p}+v_pt)^2+(y_s-y_0)^2)^{\frac{1}{2}}}] = \nonumber \\
=-cR \int d\alpha \frac{v_p q cos(\alpha)}{\sqrt{(d-r(sin(\alpha))-v_p t)^2+(H-r\cos(\alpha))^2}}+ \nonumber \\
- \int_{t0}^{t}dt' \int_{0}^{2\pi} d\alpha \frac{q^2 (2 r \sin (\alpha ) (H-r \cos (\alpha ))-2 r \cos (\alpha ) (-r \sin (\alpha
   )+d-t' v_p))}{2 \left((-r \sin (\alpha )+d-t' v_p)^2+(H-r \cos (\alpha
   ))^2\right)^{3/2}}+ \nonumber \\
- \frac{v_p q \mu_0}{4\pi}\int_{y1=-R}^{y_{2s}=-y_{1s}=R} dy_s [\frac{\sqrt{R^2-y_s^2}-x_{0p}-v_pt}{y_s-y_0} \frac{1}{((\sqrt{R^2-y_s^2}-x_{p0}-v_pt)^2+(y_s-y_0)^2)^{\frac{1}{2}}} \nonumber \\
+\frac{\sqrt{R^2-y_s^2}+x_{0p}+v_pt}{y_s-y_0} \frac{1}{((\sqrt{R^2-y_s^2}+x_{0p}+v_pt)^2+(y_s-y_0)^2)^{\frac{1}{2}}}]=\pi R^2 I(t)
%+R^2-y_s^2}-x_{0p}-v_pt}{y_s-y_0} \frac{1}{((\sqrt{R^2-y_s^2}-x_{p0}-v_pt)^2+(y_s-y_0)^2)^{\frac{1}{2}}} \nonumber \\
%+\frac{\sqrt{R^2-y_s^2}+x_{0p}+v_pt}{y_s-y_0} \frac{1}{((\sqrt{R^2-y_s^2}+x_{0p}+v_pt)^2+(y_s-y_0)^2)^{\frac{1}{2}}}]
\end{eqnarray}
\section{Bogoliubov-de Gennes formalism in modeling of robust FIJJs}
Let us concentrate on the case depicted in Fig.\ref{fig1q}a when we place long ferromagnetic strip on a thin superconductor and separate both systems by insulator. %%000For the moment let us assume that Fe strip generates small magnetic field.  Let us assume that we have one dimensional superconductor with constant superconducting order parameter (SCOP). We place this superconductor in external magnetic field produced by external coil that is sufficient small so it does not change the distribution of SCOP.
In such case we need to consider the scattering problem on non-uniform vector potential generated by ferromagnetic strip placed on the top of superconducting strip so in  general case we have three non-zero components of vector potential. %%The Schrodinger equation for 1 dimensional particle is written as
%%\begin{equation}
%%  H=\frac{1}{2m}(\hat{p}_x^2+\hat{p}_y^2+\hat{p}_z^2)+V_0                                                                                                                                                                                                                                            (x)
%%\end{equation}
%.
We can use the superconducting order parameter distribution obtained from GL ($\Delta(x)=\psi(x)$) and external biasing vector potential for exact computation of quantum eigenstates of Josephson junction (or any other superconducting structure) in Bogoliubov-de Gennes approach.
Electrons and holes moving in superconductor do not have many degrees of freedom since its y and z coordinates are squeezed to small dimension and can be ignored. Therefore one obtains the following Schrodinger Hamiltonian
$\hat{p}_x=(\frac{\hbar}{i}\frac{d}{dx}-\frac{2e}{c}A_x(x))$ and $\hat{p}_y=(-\frac{2e}{c}A_y(x)), \hat{p}_z=(-\frac{2e}{c}A_z(x))$ that can be rewritten
 $ H=\frac{1}{2m}((\frac{\hbar}{i}\frac{d}{dx}-\frac{2e}{c}A_x(x))^2+(-\frac{2e}{c}A_y(x))^2+(-\frac{2e}{c}A_z(x))^2)$ and placed into                                                                                                                                                                                                                                       Bogoliubov-de Gennes (BdGe) equations of the form:
%%$$
%\begin{matrix}
%a & b \\
%c & d
%\end{matrix}
%\quad
%\begin{pmatrix}
%a & b \\
%c & d
%\end{pmatrix}
%\quad
\begin{equation}
\begin{bmatrix}
H(x) & \Delta(x) \\
\Delta^{\dag}(x) & -H^{\dag}(x)
\end{bmatrix}
\begin{bmatrix}
u_n(x)  \\
v_n(x)  \\
\end{bmatrix}
=
\epsilon_n
\begin{bmatrix}
u_n(x)  \\
v_n(x)  \\
\end{bmatrix}
,
\Delta(x)=V \sum_n u_n(x)v_n^{\dag}(x)(1-2f(\epsilon_n)),
\label{eq1}
\end{equation}
%\quad
%\begin{vmatrix}
%a & b \\
%c & d
%\end{vmatrix}
%\quad
%\begin{Vmatrix}
%a & b \\
%c & d
%\end{Vmatrix}
%%%%%$$
where $u_n(x)$, $v_n(x)$ are wavefunctions of electron and hole and $\Delta$ is superconducting order parameter and $\epsilon_n$ are eigenenergies of BdGe equations with Fermi-Dirac distribution function denoted as $f(\epsilon_n)$ and pairing constant V.
%and hence
%\begin{equation}
%  H=(\frac{\hbar^2}{2m}\frac{d^2}{dx^2}-\frac{e}{mc}A_x(x)\frac{\hbar}{i}\frac{d}{dx})+i(\frac{e}{mc}\hbar\frac{d}{dx}A_x(x))+(\frac{2e}{c})^2((A_x(x))^2+(A_y(x))^2+A_z(x)^2)
%\end{equation}
%what can be written in the form
%\begin{equation}
%  H=(\frac{\hbar^2}{2m}\frac{d^2}{dx^2}-\frac{e}{mc}A_x(x)\frac{\hbar}{i}\frac{d}{dx})+i(\frac{e}{mc}\hbar\frac{d}{dx}A_x(x))+V_1(x)
%\end{equation}
%and
%\begin{equation}
%  V_1(x)=(\frac{2e}{c})^2((A_x(x))^2+(A_y(x))^2+A_z(x)^2)
%\end{equation}
%The operator H occurs in given form in BdGe formalism and in Ginzburg-Landau formalism.
Self-consistency for vector potential as a function of electric current distribution in the system requires that $\vec{A}(x) \propto \int dx' \vec{j}(x')/|x-x'|$ has to be fullfilled.
Simple distrubution of $A_z$ vector potential component is the case $A_z(x,y=0) \propto \frac{I_0}{(a_0^2+x^2)^{\frac{1}{2}}}$ when one uses the one polarizing cable with z non-zero current component $I_0$ and constant $a_0$.
%In general the vector potential is given as $\vec{A}(x) \approx k_1 \int dx' \vec{j}(x')/|x-x'|$.
%%The easiest calculation is determination of $A_z$ vector potential component in reference to Fig.1a that is of the form
%We can write
%%Here is the text of your introduction.
%%\begin{equation}
%%    \label{simple_equation}
%%    $A_z(x,0) = \frac{I_0}{(a_0^2+x^2)^{\frac{1}{2}}}$
%%\end{equation}
%.
%%Such situation is the most simple and it is the case of single current source generating Az component of vector potential.
%\section{Simulation methodology }
One of the constants of motion is the conservation of electric charge that introduces many constrains in quasi-one dimensional case.
%%%Certain fact that should be used in modeling quasi-one dimensional superconducting structures is electric current conservation.
Since both Bogoliubov-de Gennes model and Ginzburg-Landau (GL) can be derived from BCS theory it is easier and justified to start from GL.
In Ginzburg-Landau model conservation of electric charge essentially means that $constant=I=-c_1 A_x(x,y=0)|\psi(x,y=0)|^2$ with constant $c_1$ and SCOP $\psi$ that refers to physical situation from Fig.\ref{fig1q}a and Fig.\ref{fig1q}b.
Having the distribution of vector potential from GL we can place into BdGe equations as initial condition for computations and obtain new superconducting order parameter and new vector potential after certain number of iterations.
From computational point of view usage of BdGe is equivalent to eigenvalue problem for eigenergy and eigenstate of given Bogoliubov-de Gennes matrix from equation \ref{eq1}. In general case we shall consider time-dependent case and Zeeman splitting in equation \ref{eq1}.
%%\begin{equation}
%%\Delta(x+s \Delta x)=V \sum_n u_n(x+s \Delta x)v_n^{\dag}(x+s \Delta x)(1-2f(\epsilon_n))
%%\end{equation}

%%%New superconducting order parameter will bring new vector potential in superconducting cable. It can be relatively simply determined. Let us assume the case of Fig.1.
%%%%\begin{equation}
%%%%const=I=  -c_1 A_x(x) \sum_n (f(\epsilon_n) |u_n(x)|^2 + (1-f(\epsilon_n)) |v_n(x)|^2)
%%%%\end{equation}
%%%that brings the expression for vector potential for every point of our lattice under consideration $A_x(x+s\Delta x)$
%%%\begin{equation}
%%%\frac{I}{-c1 \sum_n (f(\epsilon_n) |u_n(x+s\Delta x)|^2 + (1-f(\epsilon_n)) |v_n(x+s\Delta x)|^2)}=A_x(x+s\Delta x)
%%%\end{equation}
By repeating this procedure many times one finds superconducting order parameter and vector potential in self-consistent way when results from previous iteration step are nearly the same as the results from current iteration step from numerical point of view.
Because of vector potential self-consistency across the whole system we obtain coupled BdGe equations that describe propagation of wavepackets in each designated superconducting wire.
In our considerations we assume that distribution of current is unchanged and fixed in polarizing cables generating magnetic field that affect and bias cables propagating BdGe wavepackets.
One can use also superconductors for biasing cables and in such case quasi-one dimensional GL equations can describe this situation together with BdGe equations related to cables when one considers the field induced Josephson junctions.
Obviously superconducting polarizing cables should have critical current much higher than critical current of cables described by BdGe wavepackets.
%It will be suggested later but we can have N+M (N are assymptotic superconductors and M are closed superconductors) interacting BdGe systems as depicted in Fig.2d. %\ref{case4}.

%%%Write your subsection text here.
\section{Future experiments and outlook}
\par
Future experiments can be designed based on the concept presented in Fig.\ref{fig2q}d. Instead of conducting very large number experiments referring to each possible case of polarizing and BdGe cable configuration one can built the system where topology of network can be controlled in electrical way. It will limit certain number of configurations and point main Cartesian directions.  In such case we have to consider two lattices (\emph{BdGe cables} + \emph{polarizing cables}) in close proximity and such that they are embedded one into each other.
One lattice will generate complex pattern of magnetic field and this will be the network of cable polarizes and magnetic field entangler (MFE). Another network will be the network of thin superconductors propagating BdGe wavepackets that we name simply BdGe cables. \par
First approach is presented in Fig.\ref{fig3q}. when one uses the network of normal cables as cable polarizes controlled by semiconductor N-P-N or P-N-P transistors. In case of BdGe cables we need to use superconducting version of transistor that is Josephson junction
with controlled channel by external voltage as presented by \cite{JJtransistor}. Such approach is very difficult in implementation since dissipative current flowing via polarizing cables will generate heat that will affect network of superconducting cables even in case of intensive cooling. What is more voltages involved in transistor operation are 3 orders of magnitude higher than voltages necessary to destroy superconducting state and bring it to normal state what requires excellent insulators.
The necessity of maintenance of  high electric insulation and cooling makes such architecture rather unrealistic in real implementation. \par
It is rather desirable to use only superconducting cables as cable polarizes. The control of their topology can be obtained by destroying locally the superconducting state with usage of another cable with high magnetic field as it is depicted in Fig.\ref{fig3q},
and in more detailed in 2 dimensions in Fig.\ref{fig4qa},\ref{fig4qb},\ref{fig4qd} and in 3 dimensions in Fig.\ref{fig4qc}. In such case we need 4 networks of cables and this suggests the usage both of low and high temperature superconductors. Two of them can be used as the cables 'cutting' other cables locally by use of generated magnetic field. The magnetic field cutting scheme is depicted in Fig.\ref{fig3q}. Since 'cutting cable' need to have higher critical current or $T_c$ than cable being cut it is recommendable to use 4 types of superconductors with different critical temperatures. Two networks doing cutting can be interpreted as the signalling networks and they will conduct the electric current of the highest intensity. In order to minimize the magnetic field coming from this current (that is parasitic magnetic field) one needs to use signalling cables in such way that cables with opposite current flow will be
kept at minimum distance in accordance to principle that dipole generates smaller value and quicker decaying field than the case of single monopole (concept also known broadly among electric engineers as twisted cables generating minimal magnetic field).
What is more one should use two loops of cutting cables with opposite flow of current as it is pointed in left Fig.\ref{fig3q} that minimizes the leakage of magnetic flux. In more detailed the proposed scheme of topology of signalling cables in relation to polarizng and BdGe cables is depicted in Fig.\ref{fig4q} (in two dimensions) and it can be generalized quite easily to 3 dimensions as it is presented in Fig.\ref{fig5q}. In general case we can engineer network of N inputs/outputs with BdGe wavepackets and M closed superconducting cables propagating BdGe wavepackets (blue cable from Fig.\ref{fig2q}d)
that are subjected to the network of $N_1$ polarizing cables of arbitrary topology. In such case one expects to obtain the network of coupling field induced Josephson junctions (shown in Fig.\ref{fig1q}a) that have continuously tuned electric parameters (for example shape of washboard potential, critical Josephson current etc.). This should bring the possibility of creating network of Josephson junctions with wide class of complex washboard potentials. Consequently this should widen class of possible schemes of implementations of classical and quantum computer in superconducting environments. What is more presented structure in Fig.\ref{fig2q}d. and having experimental implementation as subfigures of Fig.\ref{fig4q} and \ref{fig5q} can be used as physical system with continuously tuned microscopic scattering matrix controlled by outside macroscopic signals (current fed to magnetic field polarizers and cutting cables).
The bigger is 2 or 3 dimensional network the bigger is the signalling box collecting all signalling cables that is schematically depicted in left Fig.\ref{fig4qb} and in Fig.\ref{fig4qd} . %%left and right Fig.\ref{fig5q}.
\par
\begin{figure}[hbt!]
    %%\centering
    %% \includegraphics[width=4.0in]{new_design11Qp5.png}
%%\caption{Scheme of interface between cryogenic environment and room temperature electronics. $S_1$,.. $S_n$ voltages control system biasing conditions.   } %%%Strong violet (BdGe waveguides) and }
%%\end{figure}
%%\begin{figure}
    \centering
     %%\includegraphics[width=4.0in]{new_paradigm1.png}
%%\caption{Topology of cables cutting another cables by means of magnetic field. This system is 2 dimensional version of physical system from Fig.2d.  } %%%Strong violet (BdGe waveguides) and }
 %%%%    \includegraphics[scale=0.2]{3d_robust_rFIJJ1.png} %%{new_concept_3dim.png} %[width=4.0in]
 %%%%    \caption{Generalization of previous figure to 3D.}
%%\end{figure}
%% \begin{figure}
     \includegraphics[scale=0.3]{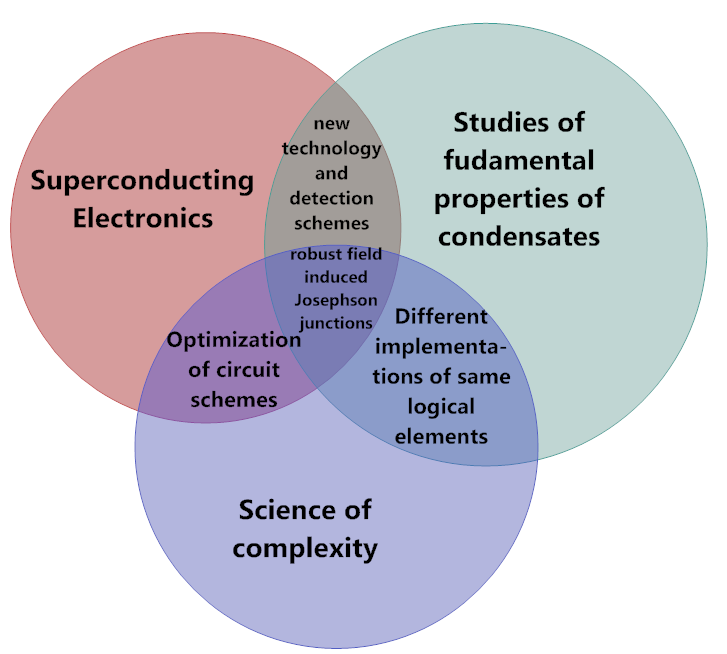} %%%{circles.png}
     \caption{Overlap of different scientific disciplines connected with study of robust field induced Josephson junctions.} % There is superposition of two metalattices in 3 dimensional space so they do not touch themselves.}% Both lattices have similar electric control by N-P-N
     \label{fig6q}
\end{figure}
\begin{figure}
    \centering
    ~ %add desired spacing between images, e. g. ~, \quad, \qquad, \hfill etc.
    \begin{subfigure}[b]{0.45\textwidth}
        \includegraphics[width=\textwidth]{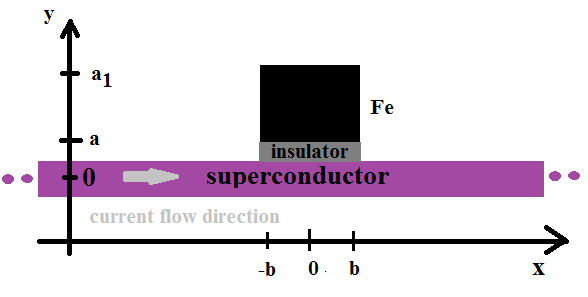}
        \caption{Physical system 1.}
        \label{fig:g3}
    \end{subfigure}
    \begin{subfigure}[b]{0.45\textwidth}
        \includegraphics[width=\textwidth]{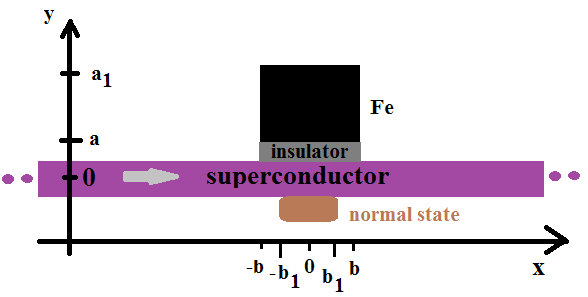}
        \caption{Physical system 2.}
        \label{fig:g4}
    \end{subfigure}
        \begin{subfigure}[b]{0.45\textwidth}
        \includegraphics[width=\textwidth]{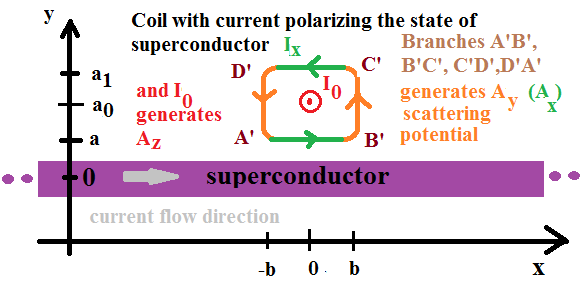}
        \caption{Scheme 1-approximation of (a)}
        \label{fig:g1}
    \end{subfigure}
    ~ %add desired spacing between images, e. g. ~, \quad, \qquad, \hfill etc.
    \begin{subfigure}[b]{0.45\textwidth}
        \includegraphics[width=\textwidth]{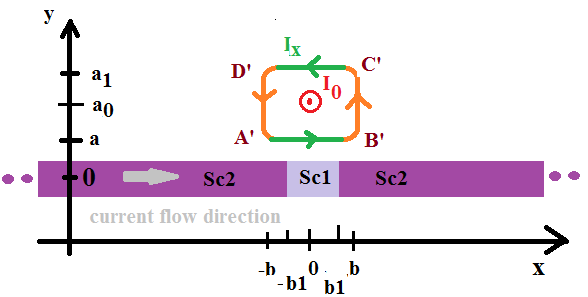}
        \caption{Scheme 2-approximation of (b)}
        \label{fig:g2}
    \end{subfigure}
    \caption{Concept of Field Induced Josephson junction in basic configurations (a or b) with approximation schemes (c or d).}
    \label{fig1q}
%%\end{figure}
%%\begin{figure}
    \centering
    \begin{subfigure}[b]{0.35\textwidth}
        \includegraphics[width=\textwidth]{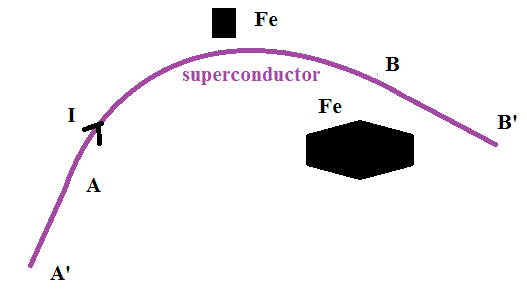}
        \caption{Deformed superconducting cable in 2D among many ferromagnets.}
        \label{fig2qa}
    \end{subfigure}
    ~ %add desired spacing between images, e. g. ~, \quad, \qquad, \hfill etc.
    \begin{subfigure}[b]{0.35\textwidth}
        \includegraphics[width=\textwidth]{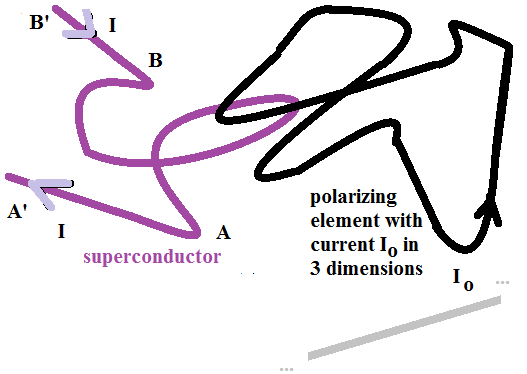}
        \caption{Deformed superconducing and polarizing cables in 3D.}
        \label{fig2qb}
    \end{subfigure}
    ~ %add desired spacing between images, e. g. ~, \quad, \qquad, \hfill etc.
    \begin{subfigure}[b]{0.35\textwidth}
        \includegraphics[width=\textwidth]{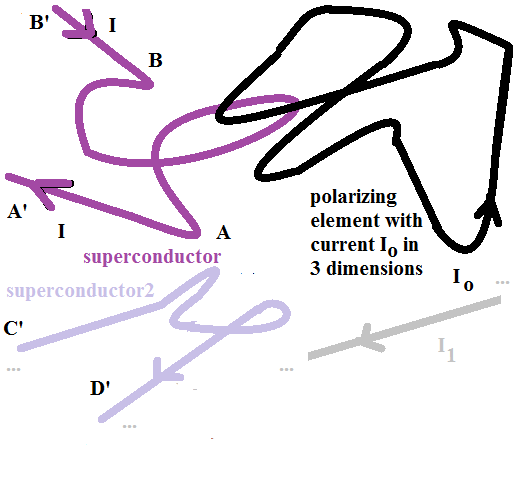}
        \caption{Two robust FIJJs interacting \\ in inductive way in the field of \\ polarizing cables.}
        \label{fig2qc}
    \end{subfigure}
    \begin{subfigure}[b]{0.35\textwidth}
        \includegraphics[width=\textwidth]{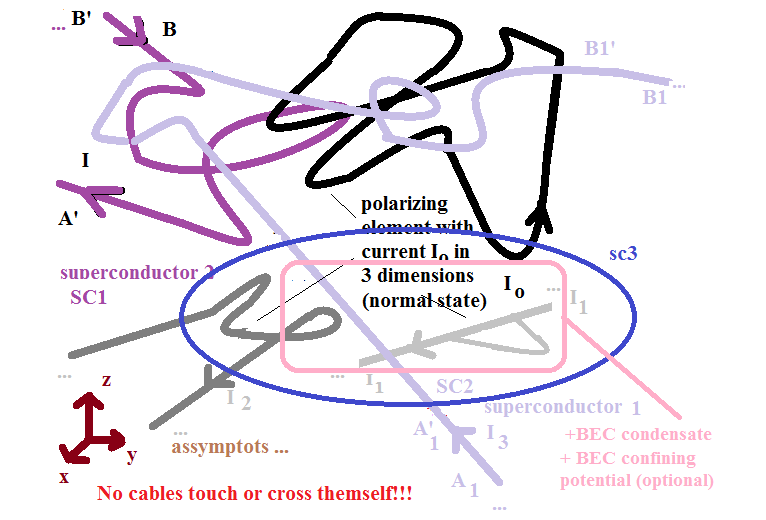}
        \caption{Robust FIJJs (rFIJJs) interacting in complex electromagnetic environment-most general case with optional existence of BEC condensate.}
        \label{fig2qd}
    \end{subfigure}
    \caption{Stages of generalization of concept of FIJJ (Field Induced Josephson junction).}
    \label{fig2q}
\end{figure}
\begin{figure}
    \centering
     \includegraphics[width=6.0in]{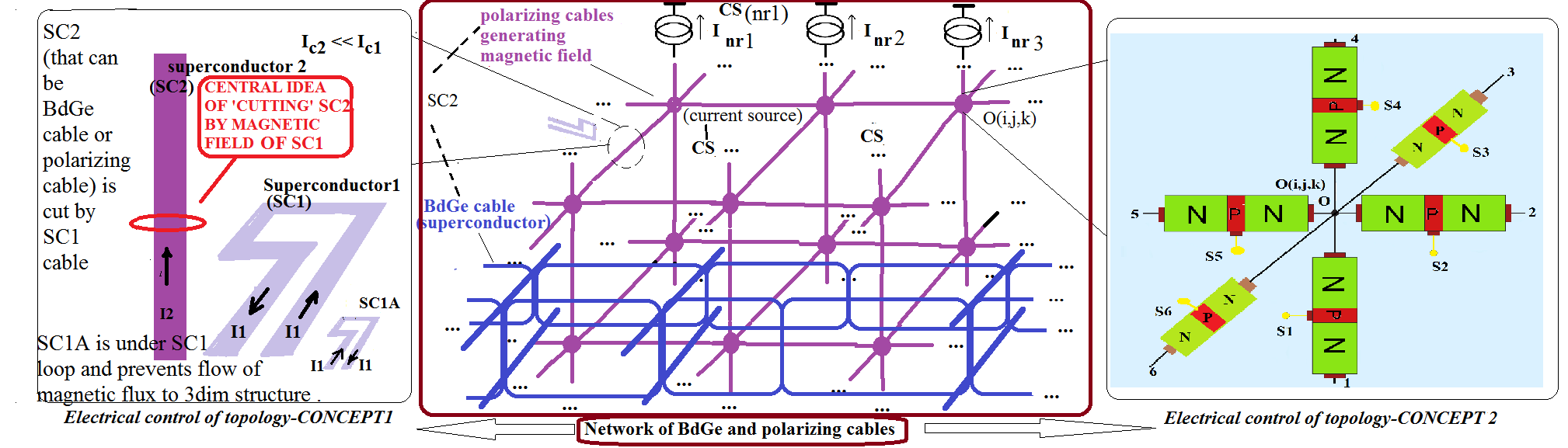} %%{magnetic_field_entanglerQ211QQ311Q.png} %%%%%\includegraphics[scale=0.3]{magnet_scissors1.png} %%%211Q.png}
\caption{Prototype of magnetic field entangler (MFE) embedded in superconducting cable environment and  connected to current sources (CS). Each node of MFE can be controlled by 6 semiconductor transistors so one can promote certain directions of current flow and hence control
topology of MFE. Same concept but basing on 6 Josephson junction transistors \cite{JJtransistor} can be assigned to BdGe cable nodes. }
 \label{fig3q}
 %%\end{figure}
 %%\begin{figure}
 \centering
 \begin{subfigure}[b]{0.45\textwidth}
    \includegraphics[scale=0.4]{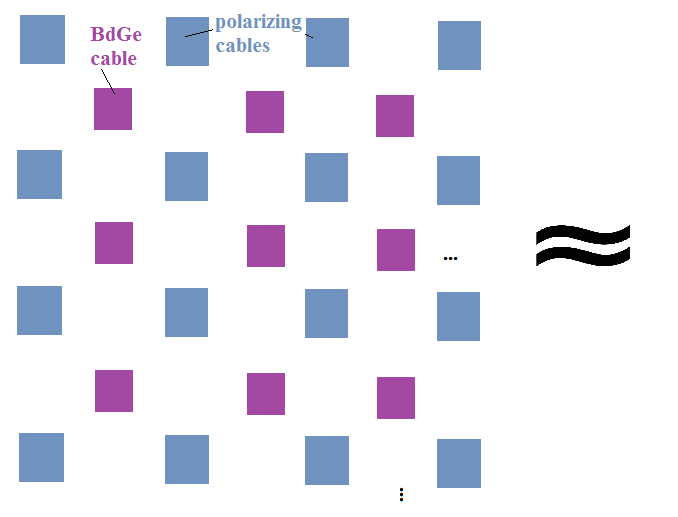}
    \caption{2 dim BdGe cables and \newline polarizing cable lattice}
    \label{fig4qa}
 \end{subfigure}
 \begin{subfigure}[b]{0.45\textwidth}
    \includegraphics[scale=0.18]{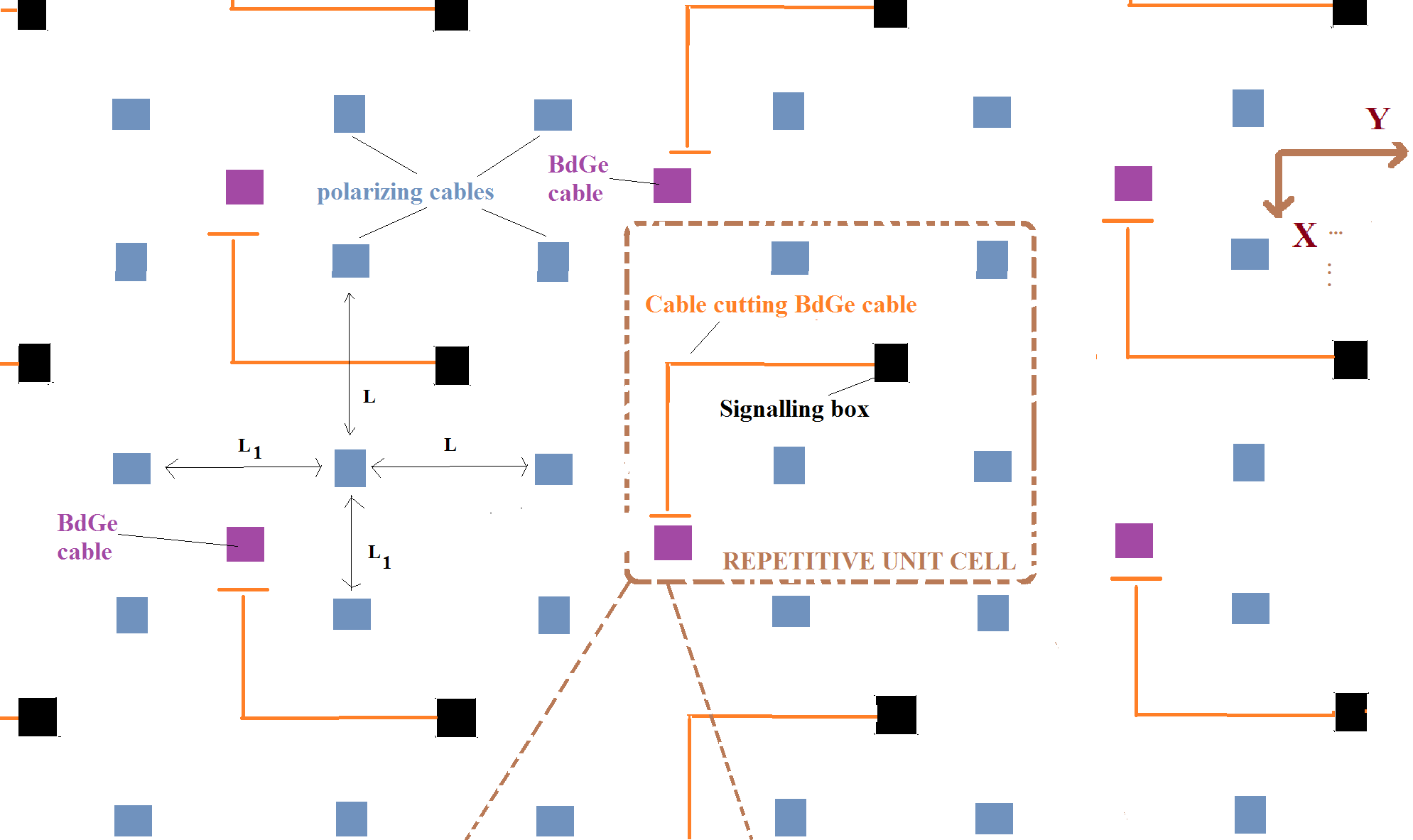}
    \caption{2 dim BdGe cable lattice + polarizing cable lattice with cutting cables and signalling boxes}
    \label{fig4qb}
 \end{subfigure}
    %% \caption{[Left]: 2 dimensional BdGe and polarizing cable network from Fig.\ref{fig2q}d. [Right]: Experimental implementation of left figure requires usage of cutting cables and signalling boxes. View of elementary cell is given below.}
       \label{fig4q}
       \begin{subfigure}[b]{0.45\textwidth}
     \includegraphics[scale=0.15]{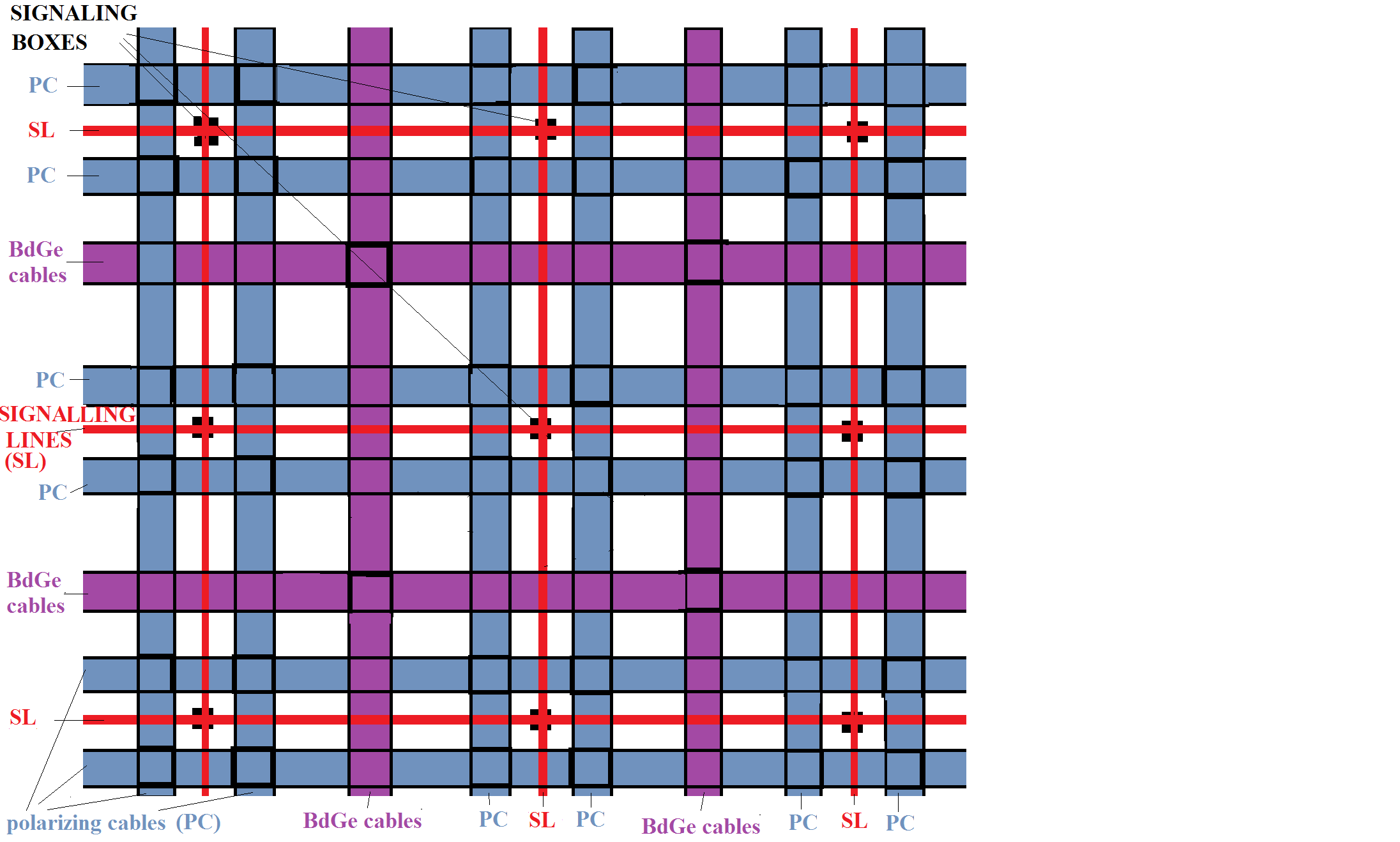}
     \caption{3 dim BdGe + polarizing cable lattice \newline + signalling cables and signalling boxes.}
     \label{fig4qc}
       \end{subfigure}
       \begin{subfigure}[b]{0.45\textwidth}
     \includegraphics[scale=0.16]{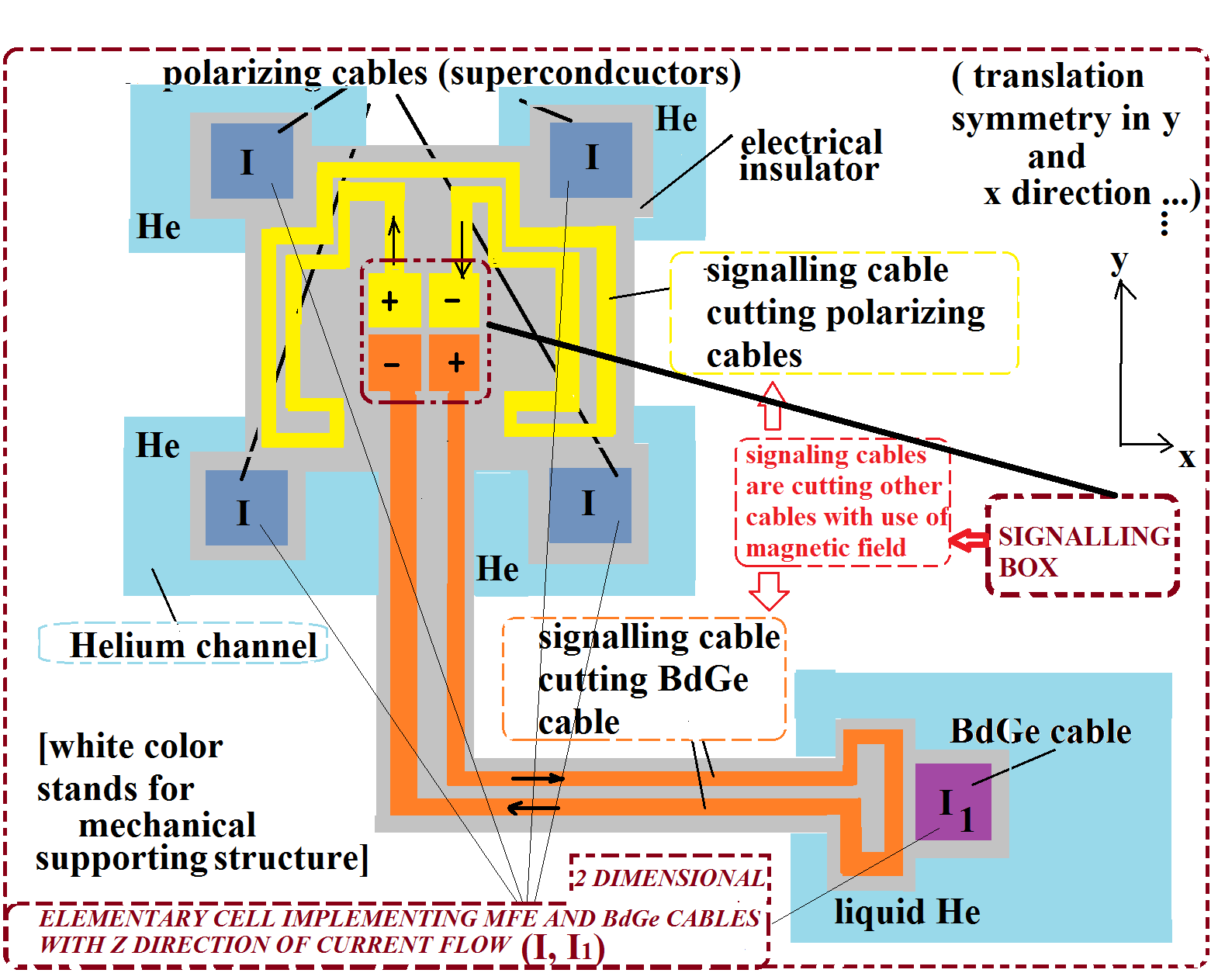}
     \caption{Detailed view of 2 dim unit cell of BdGe and polarizing cables with signalling cables and box.}
     \label{fig4qd}
     \end{subfigure}
     \caption{Detailed scheme of 2 [Fig.\ref{fig4qa} ,\ref{fig4qb}, \ref{fig4qd}] and 3 [Fig.\ref{fig4qc}.] dimensional BdGe cables with polarizing cable networks that are controlled by two signalling networks (cutting cables).}
     %%\caption{[Left]: Generalization of previous 2 dimensional network of BdGe cables and cable polarizes to 3 dimensions. [Right]: Detailed view of 2 dimensional cell implementing BdGe cables and polarizing cables together with two types of cutting cables and signalling box.}
       \label{fig5q}
 \end{figure}

In order to simulate the behaviour of the system from Fig.\ref{fig2q}d. it seems to be sufficient to use quasi-one dimensional BdGe wavepacket for signals propagating wavepackets and to use quasi-one dimensional GL
model for other cables. The self-consistency relation for vector potential as generated by all current sources shall be included in the calculations as well as self-consistency in superconducting order parameter in cables propagating BdGe wavepackets.
Therefore proposed scheme can be investigated further both in theoretical and in experimental way and is of interdisciplinary character as pointed by Fig.\ref{fig6q}. It seems to be fruitful to use concept of fitness function that is known from evolutionary algorithms (as for example in antenna design \cite{Antenna}) and science of complexity and use it to design new scheme of both classical and quantum circuits. The fitness function should take into account the correctness of logical operations and their vulnerability to error against external noise. One can conduct the artificial evolution referring to Fig.\ref{fig2q}d. in numerical simulation (artificial gene is expressed by topology of BdGe and polarizing cables) or can conductor evolution in real experiment as it is described in subfigures of Fig.\ref{fig2q},\ref{fig3q}, \ref{fig4qa} and \ref{fig4qc}. The presented scheme is expected to describe the implementation of both classical and quantum computer basing on Josephson junctions.
One expects the emergence of new schemes that are not accounted or considered as well as pointing the already existing schemes.  Such approach is already known in the context of programmable matter \cite{ProgramMatter} and suggested evolution would be example of it. The obtained new schemes can be optimized further with use of well-known methodologies.
What is more the provided scheme can also provide the platform for experimental studies of BEC condensate interacting with superconducting environment as it is depicted in Fig.\ref{fig2qd}. In order to control the interaction of BdGe cables with superconducting cables in proper way one shall use very low $T_c$
superconductors in BdGe cables.  What is more one shall use helium 3 instead of helium 4. The natural place for BEC condensate would be space in proximity to helium 3. In order to describe the system in mathematical way one needs usage of coupled Gross-Pitaevskii equations (describing BEC condensate) together with coupled BdGe and GL equations.
 In such case the basic source of concepts can be taken from \cite{hybridNori}. However they need to be readjusted for the class of systems described here. Particular expectation for the case of superconductors interacting with BEC condensate would be the emergence of robust Josephson
 junctions in BEC condensate as well as complex network of vortices. This issue is the subject of future research. Furthermore one can propose various topologies of field induced Jospephson junctions coupled to semiconductor single-electron devices (Wannier qubits) as it was indicated in \cite{SELchain} and it can lead to hybrid quantum superconducting-semiconductor computer.
 %%\par $ $
%\newline
%\newline
\section{Acknowledgment}
\par
%%%\emph{
This work is done thanks to grant JSPS-KAKENHI:26220904. I would like to thank to A.Bednorz, P.Burset, B.Lu and Y.Tanaka for discussions and remarks on presented topic.
%%}

%\begin{figure}
%    \centering
%\begin{subfigure}[b]{0.25\textwidth}
%\includegraphics[scale=0.3]{general_case2.png}
% \label{fig1:g11}
%\caption(Deformed sc cable}
%\end{subfigure}
%\begin{subfigure}[b]{0.25\textwidth}
%\includegraphics[scale=0.3]{general_case3.png}
% \label{fig1:g12}
%\caption(Deformed sc cable}
%\end{subfigure}
%\begin{subfigure}[b]{0.25\textwidth}
%\includegraphics[scale=0.3]{general_case3d.png}
% \label{fig1:g13}
%\caption(Deformed sc cable}
%\end{subfigure}
%\begin{subfigure}[b]{0.25\textwidth}
%\includegraphics[scale=0.3]{general_case3d1.png}
% \label{fig1:g14}
%\caption(Deformed sc cable}
%\end{subfigure}
%\end{figure}
%\begin{subfigure}[b]{0.25\textwidth}
%\includegraphics[scale=0.3]{general_case3.png}
%\caption(Deformed all cables}
%\end{subfigure}
%\begin{subfigure}[b]{0.25\textwidth}
%\includegraphics[scale=0.3]{general_case3d.png} %%%\includegraphics[scale=0.3]{general_case3d1.png}
%\caption{2FIJJS interacting}
%\end{subfigure}
%\end{figure}
%\begin{figure}
%    \centering
%     %%\includegraphics[width=4.0in]{general_case3d2Q.png}
%     \includegraphics[scale=0.3]{general_case3d1.png}\includegraphics[scale=0.3]{general_case3d2Q.png}
%\caption{Case of two interacting superconducting field induced Josephson junction with presence of externally closed superconducting loops. Cable might form many possible knots.  Conceptual stages leading to generalization of system of coupled field induced Josephson junctions.}
%\label{case4}
%\end{figure}

\begin{figure}[hbt!]
    \centering
     \includegraphics[width=4.0in]{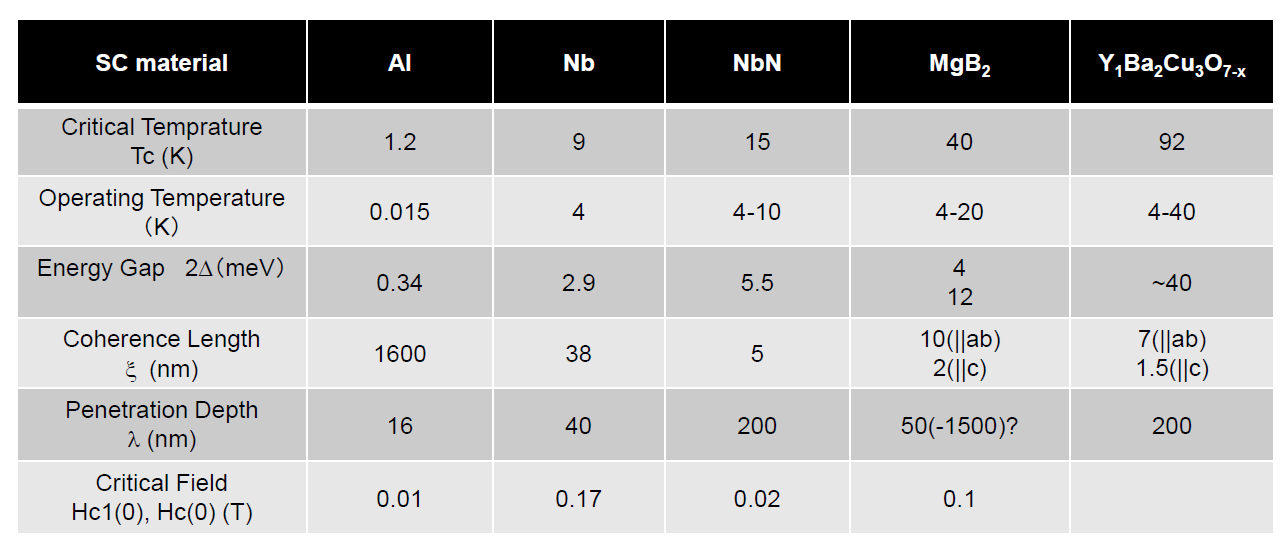} \\ %%{magnetic_field_entanglerQ211QQ311Q.png} %%%%%\includegraphics[scale=0.3]{magnet_scissors1.png} %%%211Q.png}
     \includegraphics[width=4.0in]{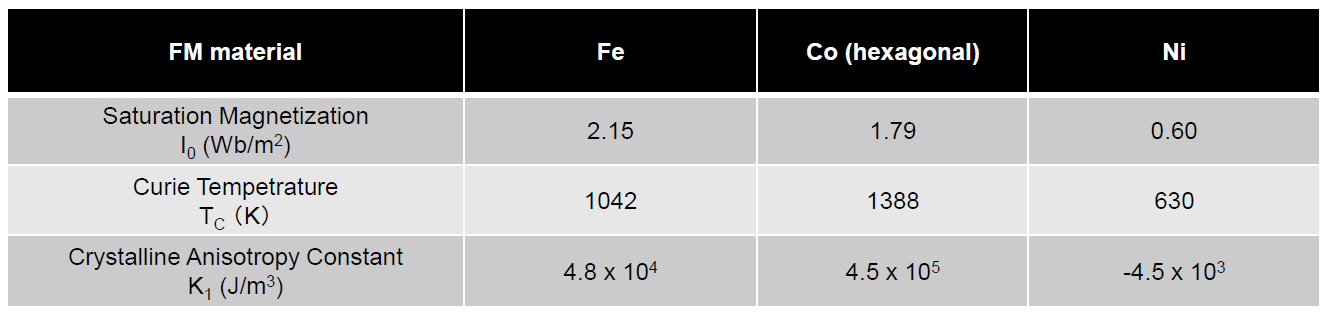} \\
Table 1: Example of superconductor and ferromagnetic material properties that are relevant in robust field induced Josephson junction design. Superconductor nanowires in dimensions not exploited in design can have the size between one and two superconducting coherence length. The dimensions of the whole closed superconducting loop is at least 10 coherence lengths.
\label{T1}
\end{figure}


\newpage



\begin{thebibliography}{99}
\parskip -5pt
%
\bibitem{Josephson}
B.D.Josephson, Possible new effects in superconductive tunnelling, %Physics Letters 1, 251,
%1962
%H.~Myoren, Y.~Wakimizu, and S.~Takada,
{\itshape Physics Letters,}
vol.~1, no.~251, 1962.

\bibitem{Likharev}
%S.~Yorozu, Y.~Kameda, H.~Terai, A.~Fujimaki, T.~Yamada, and S.~Tahara,
%{\itshape Physica C,}
%vol.~378--381, pp.~1471--1474, 2002.
K.K.Likharev et al.,RSFQ logic/memory family, Rapid Single Flux Quantum Logic, A new Josephson-junction technology for sub-terahertz clock-frequency digital systems,
{\itshape IEEE Transaction on Applied Superconductivity,}, vol.~1, no.~1, 1991

\bibitem{SQFMicroprocessor}
  Y.Yamanshi et al, Design and Implementation of a Pipelined Bit-Serial SFQ Microprocessor, CORE 1 $\beta$,{\itshape IEEE Transaction on Applied Superconductivity,} vol.~17, no.~2, 2007

\bibitem{Mukhanov}
O.A.Mukhanov, Energy-efficient single flux quantum technology,{\itshape IEEE Transaction on Applied Superconductivity,}, vol.~21, no.~3, 2011,

\bibitem{Nori}
J.Q.You, F.Nori, Superconducting circuits and quantum information, {\itshape Physics today}, vol.~48, no.~52, 2005

\bibitem{Clinton}
 T.W.Clinton et al.,Nonvolatile switchable Josephson junctions, {\itshape Journal of Applied Physics,} vol.~85, no.~1637, 1999

\bibitem{Maeda}
 A.Maeda, L.Gomez, Experimental Studies to Realize Josephson Junctions and Qubits in
Cuprate and Fe-based Superconductors, {\itshape Journal of Superconductivity and Novel magnetism }, vol.~23, no.~5,
2012

\bibitem{Pomorski}
K.Pomorski, P.Prokopow, Possible existence of field induced Josephson junctions, {\itshape Physica Status Solidi B}, 2012 + K.Pomorski PhD thesis:Physical description of unconventional Josephson junctions, Jagiellonian University, 2015


\bibitem{JJtransistor}
Tatsushi Akazaki et al., A Josephson field effect transistor using an InAs-inserted-channel $In_{0.52}Al_{0.48}As/In_{0.53}Ga_{0.47}As$ inverted modulation-doped structure,  {\itshape Applied Physics Letters,} vol.~68, no.~3, 1996

\bibitem{Nazarov}
J.E.Mooij et al. , Superconducting nanowires as quantum phase-slip junctions,  {\itshape Nature Physics} , vol.~2, 2006

\bibitem{BTK}
G.E.Blonder et al, Transition from metalic to tunneling regimes in superconductor constrictions:excess current, charge imbalance and supercurrent conversion, {\itshape Physical Review B},  vol.~25, no.~7, 1982

\bibitem{Antenna}
Y.Li, Simulation-based evolutionary method in antenna design optimization, {\itshape Mathematical and computer modeling},  vol.~51, 2010

\bibitem{ProgramMatter}
T.Toffoli, Programmable matter methods, {\itshape Future generation computer systems},vol.~16,1999

\bibitem{hybridNori}
Ze-Liang Xiang et al. , Hybrid quantum systems: Superconducting circuits interacting with other quantum systems, {\itshape Reviews of modern physics}, vol.~85, no.~2, 2013

\bibitem{Compel}
 K.D.Pomorski, H.Akaikea, A.Fujimaki, K.Rusek, Relaxation method in description of RAM memory cell in RSFQ computer,  {\itshape The international journal for computation and mathematics in electrical and electronic engineering} , {\itshape Compel}, vol.~38, no.1, 2019

\bibitem{SELchain}
 K.D.Pomorski, P.Peczkowski, R.B.Staszewski, Analytical solutions for N interacting electron system confined in graph of coupled electrostatic semiconductor and superconducting quantum dots in tight-binding model, {\itshape Cryogenics}, vol.~109, Page: 103117, 2020

%\bibitem{FaDu89}
%E.~S.~Fang and T.~Van~Duzer,
%Ext. Abs. 1989 International Superconductivity Electronics Conference (ISEC'89),
%pp.~407--410, 1989.
%
\end{thebibliography}
\end{document}